\documentclass[a4paper,amsmath,amssymb,color,nofootinbib,preprint,tightenlines]{revtex4-2}
\usepackage{graphicx}
\usepackage[dvipsnames]{xcolor}
\usepackage[mathscr]{euscript}
\usepackage{slashed}
\usepackage{tablefootnote}
\usepackage[normalem]{ulem}
\usepackage[colorlinks=true,pdfstartview=FitV,citecolor=blue,linkcolor=purple,urlcolor=Sepia]{hyperref}
\usepackage{orcidlink}

\begin{document}
\baselineskip=17pt \parskip=3pt

\title{\boldmath\textit{CP} violation in the hyperon decays $\mathit\Sigma\to N\pi$}

\author{Xiao-Gang He\orcidlink{0000-0001-7059-6311}$,^{1,2,}$\footnote{\email{hexg@sjtu.edu.cn}}
Xiao-Dong Ma\orcidlink{0000-0001-7207-7793}$,^{3,4,}$\footnote{\email{maxid@scnu.edu.cn}}
Jusak Tandean\orcidlink{0000-0001-7581-2351}$,^{1,2,}$\footnote{\email{jtandean@yahoo.com}}
and German Valencia\orcidlink{0000-0001-6600-1290}$^{5,}$\footnote{\email{german.valencia@monash.edu}} \vspace{1ex} \\ \it
$^1$State Key Laboratory of Dark Matter Physics, Tsung-Dao Lee Institute and School of Physics and Astronomy,
Shanghai Jiao Tong University, Shanghai 201210, China\vspace{1ex} \\
$^2$Shanghai Key Laboratory for Particle Physics and Cosmology,
Key Laboratory for Particle Astrophysics and Cosmology (MOE),
School of Physics and Astronomy, Shanghai Jiao Tong University, Shanghai 201210, China\vspace{1ex} \\
$^3$State Key Laboratory of Nuclear Physics and Technology, Institute of Quantum Matter, South China Normal University, Guangzhou 510006, China\vspace{1ex} \\
$^4$Guangdong Basic Research Center of Excellence for Structure and Fundamental Interactions of Matter, Guangdong Provincial Key Laboratory of Nuclear Science, Guangzhou 510006, China\vspace{1ex} \\
$^5$School of Physics and Astronomy, Monash University, Wellington Road, Clayton, Victoria 3800, Australia
\vspace{2em} \\ \rm
Abstract
\vspace{8pt} \\
\begin{minipage}{0.99\textwidth} \baselineskip=17pt \parindent=3ex \small
The study of $CP$ violation in hyperon transitions has a long history.
In the early 2000s the HyperCP experiment made a major effort to seek $CP$-odd signals in the decay sequence $\mathit\Xi^-\to\mathit\Lambda \pi^-$ and $\mathit\Lambda\to p\pi^-$, which motivated more searches.
Most recently the BESIII and LHCb Collaborations have acquired or improved the upper bounds on $CP$ violation in a variety of hyperon nonleptonic processes, including $\mathit\Sigma^+\to n\pi^+$ and $\mathit\Sigma^+\to p\pi^0$.
These measurements have not reached the standard-model level yet, but have stimulated a renewed interest in $CP$-violating new physics in strange-quark decay beyond what is constrained by the parameters $\varepsilon$ and $\varepsilon^\prime$ from the kaon sector.
In this paper, after updating the standard-model expectations for $CP$-odd observables in the modes $\mathit\Sigma^\pm\to N\pi$, we revisit new-physics scenarios that could enhance the corresponding quantities in $\mathit\Lambda\to N\pi$ and $\mathit\Xi\to\Lambda\pi$ and apply them to the $\mathit\Sigma^\pm$ modes.
We find that the $CP$ asymmetries in the latter can be significantly increased over the standard-model expectations, at levels which may be tested in the ongoing BESIII  experiment and in future endeavors such as PANDA and the Super Tau Charm Facility.
\end{minipage}}
\maketitle

\newpage

{\bf\small\hypersetup{linkcolor=black}\tableofcontents}

\newpage

\section{Introduction\label{intro}}

Investigations concerning the phenomenon of $CP$ violation---the breaking of the combination of charge-conjugation ($C$) and parity ($P$) symmetries---aim to achieve a comprehensive understanding about its nature in order to ascertain whether or not its origin lies exclusively within the standard model (SM).
This entails searching for signs of $CP$ violation in as many processes as possible.
Although it has been established in the kaon, bottomed-meson, and charmed-meson systems since a while ago~\cite{ParticleDataGroup:2024cfk}, only very recently has it been observed among baryons, particularly in the decay of bottomed-baryon $\mathit\Lambda_b^0$ measured by the LHCb experiment~\cite{LHCb:2025ray}.
This provides an additional incentive to look for $CP$ violation in other baryon systems.

The quest for $CP$ violation in the decays of light hyperons has been performed over the years by various collaborations~\cite{R608:1985fmh,Barnes:1987vc,DM2:1988ppi,Barnes:1996si,E756:2000rge,HyperCP:2004zvh,HyperCP:2006ktj,BES:2009zvb,BESIII:2018cnd,BESIII:2022qax,BESIII:2021ypr,BESIII:2022lsz,BESIII:2022lsz,BESIII:2023lkg,BESIII:2023drj,BESIII:2023jhj,BESIII:2024nif,BESIII:2020fqg,BESIII:2023sgt,BESIII:2025jxt,Belle:2022uod,LHCb:2024tnq}, most recently by BESIII~\cite{BESIII:2018cnd,BESIII:2021ypr,BESIII:2022qax,BESIII:2022lsz,BESIII:2023lkg,BESIII:2023drj,BESIII:2023jhj,BESIII:2024nif,BESIII:2020fqg,BESIII:2023sgt,BESIII:2025jxt}, Belle~\cite{Belle:2022uod}, and LHCb~\cite{LHCb:2024tnq}, but with negative results so far.
The majority of these efforts concentrated on the modes \,$\mathit\Lambda\to p\pi^-,n\pi^0$\, and \,$\mathit\Xi^{-,0}\to\mathit\Lambda\pi^{-,0}$,\, plus their antiparticle counterparts, and the main $CP$-odd observable probed was \,$\hat A\equiv(\alpha + \overline\alpha)/(\alpha - \overline\alpha)$,\, with $\alpha$ being one of the decay asymmetry parameters and $\overline\alpha$ that of the antihyperon decay, as \,$\overline\alpha=-\alpha$\, if $CP$ is conserved.
The corresponding analyses within and beyond the SM have also been carried out~\cite{Pais:1959zza,Iqbal:1989vc,He:1991pf,Salone:2022lpt,He:2025jfc,Guo:2025bfn,Brown:1983wd,Chau:1983ei,Tandean:2002vy,Donoghue:1986hh,Donoghue:1985ww,He:1995na,Deshpande:1994vp,Chang:1994wk,He:1995na,He:1999bv,Chen:2001cv,Tandean:2003fr,Tandean:2004mv}.

Searches for $CP$ violation in $\mathit\Sigma^+$-hyperon decays have been conducted by BESIII as well~\cite{BESIII:2020fqg,BESIII:2023sgt,BESIII:2025jxt}, and the latest outcomes are \cite{BESIII:2023sgt,BESIII:2025jxt}
\begin{align} \label{ASigma-data}
\hat A_{\mathit\Sigma^+\to n\pi^+}^{\rm exp} & \,=\, -0.080 \pm 0.052_{\rm stat} \pm 0.028_{\rm syst} \,,
\nonumber \\
\hat A_{\mathit\Sigma^+\to p\pi^0}^{\rm exp} & \,=\, -0.0118 \pm 0.0083_{\rm stat} \pm 0.0028_{\rm syst} \,. ~~~ ~~
\end{align}
The predictions for them and other $CP$-asymmetries of \,$\mathit\Sigma\to N\pi$\, in SM and new-physics contexts were made decades ago~\cite{Overseth:1969bxc,Brown:1983wd,Chau:1983ei,Donoghue:1986hh,Tandean:2002vy,Okubo:1958zza}.
In view of these new data and in anticipation of more results from BESIII and of future measurements in the Belle II~\cite{Belle:2022uod} and PANDA~\cite{PANDA:2020zwv} experiments and at the proposed Super Tau Charm Facility~\cite{Achasov:2023gey}, it is therefore timely to give an up-to-date theoretical treatment of the aforesaid $\mathit\Sigma$ observables.

In this paper we first deal with the SM expectations for them, employing the currently available pertinent information, and subsequently explore potentially sizable contributions to them that hail from beyond the SM, such as those which can magnify the so-called chromomagnetic-penguin interactions~\cite{Donoghue:1985ww,Donoghue:1986hh,He:1995na,Deshpande:1994vp,Chang:1994wk,He:1999bv,Chen:2001cv,Tandean:2003fr,Tandean:2004mv} or arise in a two-Higgs-doublet scenario with dark matter~\cite{He:2025jfc} motivated by recent \,$B^+\to K^+\nu\bar\nu$\, measurements~\cite{Belle-II:2023esi}.
Our study includes \,$\mathit\Sigma^-\to n\pi^-$,\, but no decay channels of $\mathit\Sigma^0$, as its width is overwhelmingly dominated by the $\mathit\Lambda\gamma$ one \cite{ParticleDataGroup:2024cfk}.

The rest of the paper is organized as follows.
In section \ref{decayamps} we begin by evaluating the amplitudes for \,$\mathit\Sigma\to N\pi$\, from their latest data.
Then we discuss the observables sensitive to $CP$ violation that are covered in this work.
In section \ref{smpredict} we estimate their values within the SM.
In section~\ref{np1} we entertain the possibility that new physics beyond the SM greatly affects them, focusing on two different scenarios.
In the first one, taking a model-independent approach, we examine specifically the impact of chromomagnetic-penguin operators (CMOs) that is amplified by new physics.
How they might influence the $CP$ asymmetries of \,$\mathit\Sigma\to N\pi$\, has not been addressed in the recent past.
The second scenario is the aforementioned dark-matter model.
After briefly describing its salient features, we look at the $CP$-violating $\mathit\Sigma$ observables that it may enlarge above the SM expectations, taking into account dark-matter and kaon constraints.
We present our conclusions in section~\ref{conclusions}.
An appendix supplies further details relevant to the $\mathit\Sigma^-$ mode, and another one contains, for completeness, a concise updated assessment of the CMO contributions to the $CP$ asymmetries in the $\mathit\Lambda$ and $\mathit\Xi$ cases.

\section{Decay amplitudes and \textit{CP}-odd observables\label{decayamps}}

\subsection{Empirical information}

The amplitude for \,$\mathit\Sigma\to N\pi$\, is of the form~\cite{ParticleDataGroup:2024cfk}
\begin{equation}
i{\cal M}_{\mathit\Sigma\to N\pi}^{} \,=\, \bar u_N^{} \big( \mathbb A_{\mathit\Sigma\to N\pi}^{} - \gamma_5^{}\, \mathbb B_{\mathit\Sigma\to N\pi}^{} \big) u_\mathit\Sigma^{} \,,
\end{equation}
where $\mathbb A$ and $\mathbb B$ are generally complex constants belonging to, respectively, the {\tt S}- and ${\tt P}$-wave components of the transition and $u_{\mathit\Sigma,N}$ denote the Dirac spinors of the baryons.
This leads to the decay asymmetry parameters $\alpha$, $\beta$, and $\gamma$ given by~\cite{ParticleDataGroup:2024cfk,Lee:1957qs}
\begin{align} \label{alphaS2Npi}
\alpha_{\mathit\Sigma\to N\pi}^{} & \,=\, \frac{2\, {\rm Re}\big(\mathbb A_{\mathit\Sigma\to N\pi}^*\, \mathbb B_{\mathit\Sigma\to N\pi}^{}\big)\, \mathscr K_{\mathit\Sigma N\pi}^{}}{|\mathbb A_{\mathit\Sigma\to N\pi}|^2 + |\mathbb B_{\mathit\Sigma\to N\pi}\, \mathscr K_{\mathit\Sigma N\pi}|^2} \,, &
\beta_{\mathit\Sigma\to N\pi}^{} & \,=\, \frac{2\, {\rm Im}\big(\mathbb A_{\mathit\Sigma\to N\pi}^*\, \mathbb B_{\mathit\Sigma\to N\pi}^{}\big)\, \mathscr K_{\mathit\Sigma N\pi}^{}}{|\mathbb A_{\mathit\Sigma\to N\pi}|^2 + |\mathbb B_{\mathit\Sigma\to N\pi}\, \mathscr K_{\mathit\Sigma N\pi}|^2} \,, ~~~
\nonumber \\
\gamma_{\mathit\Sigma\to N\pi}^{} & \,=\, \frac{|\mathbb A_{\mathit\Sigma\to N\pi}|^2 - |\mathbb B_{\mathit\Sigma\to N\pi}\, \mathscr K_{\mathit\Sigma N\pi}|^2}{|\mathbb A_{\mathit\Sigma\to N\pi}|^2 + |\mathbb B_{\mathit\Sigma\to N\pi}\, \mathscr K_{\mathit\Sigma N\pi}|^2} \,,
\end{align}
and rate $\Gamma$
\begin{align}
\Gamma_{\mathit\Sigma\to N\pi}^{} & \,=\, \frac{({\tt M}_\mathit\Sigma+{\tt M}_N)^2-{\tt M}_\pi^2}{16\pi\, {\tt M}_\mathit\Sigma^3}\, \sqrt{\lambda_{\mathit\Sigma N\pi}}\, \big(|\mathbb A_{\mathit\Sigma\to N\pi}|^2 + |\mathbb B_{\mathit\Sigma\to N\pi}\, \mathscr K_{\mathit\Sigma N\pi}|^2\big) \,, &
\end{align}
where
\begin{align} \label{K}
\mathscr K_{\mathit\Sigma N\pi}^{} & \,=\, \frac{\sqrt{({\tt M}_\mathit\Sigma-{\tt M}_N)^2-{\tt M}_\pi^2}}
{\sqrt{({\tt M}_\mathit\Sigma+{\tt M}_N)^2-{\tt M}_\pi^2}} \,, & \lambda_{\mathit\Sigma N\pi}^{} & \,=\, \big({\tt M}_\mathit\Sigma^2-{\tt M}_N^2-{\tt M}_\pi^2\big){}^2 - 4\,{\tt M}_N^2\,{\tt M}_\pi^2 \,, ~~~ ~~
\end{align}
and ${\tt M}_{\mathit\Sigma,N,\pi}$ are the observed masses of the hadrons.
The parameters in eq.\,(\ref{alphaS2Npi}) are not all independent, satisfying \,$\alpha_{\mathit\Sigma\to N\pi}^2+\beta_{\mathit\Sigma\to N\pi}^2+\gamma_{\mathit\Sigma\to N\pi}^2=1$,\, and are linked to a fourth one, $\phi_{\mathit\Sigma\to N\pi}$, by~\cite{ParticleDataGroup:2024cfk,Lee:1957qs} \,$\beta_{\mathit\Sigma\to N\pi}^{} = \sqrt{1-\alpha_{\mathit\Sigma\to N\pi}^2}\, \sin\phi_{\mathit\Sigma\to N\pi}^{}$\, and \,$\gamma_{\mathit\Sigma\to N\pi}^{} = \sqrt{1-\alpha_{\mathit\Sigma\to N\pi}^2}\, \cos\phi_{\mathit\Sigma\to N\pi}^{}$.\,

To calculate the hyperon $CP$-violating quantities, we need the isospin components of $\mathbb A$ and $\mathbb B$ from data.
Explicitly showing the strong and weak phases, $\delta$s and $\xi$s, respectively, we can express~\cite{Donoghue:1986hh,Overseth:1969bxc}\footnote{In the sign convention which we have adopted for these amplitudes, the isospin
states  $|I,I_3\rangle$  for the initial and final hadrons are
\,$|\mathit\Sigma^\pm\rangle=\mp|1,\pm1\rangle$,\,
$|p\rangle=|1/2,1/2\rangle,\,$  $|n\rangle=|1/2,-1/2\rangle,\,$
$|\pi^\pm\rangle=\mp|1,\pm1\rangle,\,$  and \,$|\pi^0\rangle= |1,0\rangle,\,$
which are consistent with the structure of the  $\varphi$
and  $B$ matrices in eq.\,(\ref{fields}).\medskip}
\begin{align} \label{AB}
\mathbb A_{\mathit\Sigma^+\to n\pi^+}^{} & = \tfrac{2}{3}\, {\tt A}_1^{}\, e^{i\xi_1^{\rm S} + i\delta_1^{\rm S}} + \tfrac{1}{3}\, {\tt A}_3^{}\, e^{i\xi_3^{\rm S} + i\delta_3^{\rm S}} \,, & \mathbb B_{\mathit\Sigma^+\to n\pi^+}^{} & = \tfrac{2}{3}\, {\tt B}_1^{}\, e^{i\xi_1^{\rm P} + i\delta_1^{\rm P}} + \tfrac{1}{3}\, {\tt B}_3^{}\, e^{ i\xi_3^{\rm P} + i\delta_3^{\rm P}} \,,
\nonumber \\
\mathbb A_{\mathit\Sigma^+\to p\pi^0}^{} & = \tfrac{\sqrt2}{3} \Big[ {\tt A}_1^{}\, e^{ i\xi_1^{\rm S} + i\delta_1^{\rm S} } - {\tt A}_3^{}\, e^{i\xi_3^{\rm S} + i\delta_3^{\rm S}} \Big] , & \mathbb B_{\mathit\Sigma^+\to p\pi^0}^{} & = \tfrac{\sqrt2}{3} \Big[ {\tt B}_1^{}\, e^{ i\xi_1^{\rm P} + i\delta_1^{\rm P} } - {\tt B}_3^{}\, e^{ i\xi_3^{\rm P} + i\delta_3^{\rm P}} \Big] ,
\nonumber \\
\mathbb A_{\mathit\Sigma^-\to n\pi^-}^{} & = \bigg[ {\tt A}_{1,3}^{}\, e^{i\xi_{13}^{\rm S}} + \sqrt{\tfrac{2}{5}}\, {\tt A}_{3,3}^{}\, e^{i\xi_{33}^{\rm S}} \bigg] e^{i\delta_3^{\rm S}} ,  & \mathbb B_{\mathit\Sigma^-\to n\pi^-}^{} & = \bigg[ {\tt B}_{1,3}^{}\, e^{i\xi_{13}^{\rm P}} + \sqrt{\tfrac{2}{5}}\, {\tt B}_{3,3}^{}\, e^{i\xi_{33}^{\rm P}} \bigg] e^{i\delta_3^{\rm P}} ,
\end{align}
where  ${\tt A}_1$, ${\tt A}_3$, ${\tt B}_1$, ${\tt B}_3$, ${\tt A}_{2\Delta I,2I_{\rm f}}$, and  ${\tt B}_{2\Delta I,2I_{\rm f}}$ are real constants, with $\Delta I$ standing for the isospin change between the initial and final states and $I_{\rm f}$ being the total isospin of the final state,
\begin{align}
{\tt A}_1^{}\, e^{i\xi_1^{\rm S}} & \,=\, {\tt A}_{1,1}^{}\, e^{i\xi_{11}^{\rm S}} + \tfrac{1}{2}\, {\tt A}_{3,1}^{}\, e^{i\xi_{31}^{\rm S}} \,, & {\tt A}_3^{}\, e^{i\xi_3^{\rm S}} & \,=\, {\tt A}_{1,3}^{}\,e^{i\xi_{13}^{\rm S}} - \sqrt{\tfrac{8}{5}}~ {\tt A}_{3,3}^{}\, e^{i\xi_{33}^{\rm S}} \,, ~~~ ~~
\nonumber \\
{\tt B}_1^{}\, e^{i\xi_1^{\rm P}} & \,=\, {\tt B}_{1,1}^{}\, e^{i\xi_{11}^{\rm P}} + \tfrac{1}{2}\, {\tt B}_{3,1}^{}\, e^{i\xi_{31}^{\rm P}} \,, & {\tt B}_3^{}\, e^{i\xi_3^{\rm P}} & \,=\, {\tt B}_{1,3}^{}\, e^{i\xi_{13}^{\rm P}} - \sqrt{\tfrac{8}{5}}~ {\tt B}_{3,3}^{}\, e^{i\xi_{33}^{\rm P}} \,,
\end{align}
and we have neglected the \,$\Delta I=5/2$\, components,\footnote{The $\mathbb A$ and $\mathbb B$ formulas for \,$\mathit\Sigma^-\to n\pi^-$\, retaining the \,$\Delta I=5/2$\, components are written down in eq.\,(\ref{AB-S}).\medskip} which is in line with the fact that the effective quark operators responsible for the $\mathit\Sigma^\pm$ decays within the SM and new-physics scenarios considered in later sections can give rise to only \,$\Delta I\le3/2$\, interactions at leading order.\footnote{The \,$\Delta I=5/2$\, contribution can be induced by isospin breaking \cite{Gardner:2000sb}, but we ignore this possibility here.\medskip}
We notice that ${\tt A}_{1,1}$ and ${\tt A}_{3,1}$ $({\tt B}_{1,1}$ and ${\tt B}_{3,1})$ are present only in ${\tt A}_1$ $({\tt B}_1)$ and consequently cannot be individually assessed from experiment.

In table \ref{data} we collect the $\mathbb A$ and $\mathbb B$ extracted from the existing empirical information on $\alpha$ and $\gamma$ for \,$\mathit\Sigma^\pm\to N\pi$\, and their branching fractions~\cite{Bangerter:1969fta,Berley:1970zf,Harris:1970kq,Bellamy:1972fa,Lipman:1973mz,Harris:1970kq,Bogert:1970kp,Hansl:1978pc,BESIII:2020fqg,BESIII:2023sgt,BESIII:2025jxt,ParticleDataGroup:2024cfk},\footnote{In table \ref{data}, for each mode the sign of $\alpha$ ($\gamma$) fixes the relative sign (size) of $\mathbb A$ and $\mathbb B\,\mathscr K$.
The relative signs of $\mathbb A$ ($\mathbb B$) among the three modes are chosen such that the \,$\Delta I=1/2$\, rule is approximately fulfilled, their overall signs being consistent with those in the literature~\cite{Bijnens:1985kj,Jenkins:1991bt,AbdEl-Hady:1999llb,Tandean:2002vy,Donoghue:2022wrw}.} after dropping all of the phases ($\delta$s and $\xi$s), which are small.
Accordingly, the isospin components as defined in eq.\,(\ref{AB}) are
\begin{align} \label{xAxB}
{\tt A}_1^{} & \,=\, \mathbb A_{\mathit\Sigma^+\to n\pi^+}^{} + \tfrac{1}{\sqrt2}\, \mathbb A_{\mathit\Sigma^+\to p\pi^0}^{} \,=\, (-2.16\pm0.01)\times10^{-7} \,,
\nonumber \\
{\tt A}_{1,3}^{} & \,=\, \tfrac{2}{3}\, \mathbb A_{\mathit\Sigma^-\to n\pi^-}^{} + \tfrac{1}{3}\, \mathbb A_{\mathit\Sigma^+\to n\pi^+}^{}
- \tfrac{\sqrt2}{3}\, \mathbb A_{\mathit\Sigma^+\to p\pi^0}^{}
\,=\, (4.39\pm0.01)\times10^{-7} \,,
\nonumber \\
{\tt A}_{3,3}^{} & \,=\, \sqrt{\tfrac{5}{18}} \Big(
\mathbb A_{\mathit\Sigma^-\to n\pi^-}^{} - \mathbb A_{\mathit\Sigma^+\to n\pi^+}^{} + \sqrt2~ \mathbb A_{\mathit\Sigma^+\to p\pi^0}^{} \Big) \,=\, (-0.19\pm0.02)\times10^{-7} \,,
\nonumber \\ \raisebox{1em}{}
{\tt B}_1^{} & \,=\, \mathbb B_{\mathit\Sigma^+\to n\pi^+}^{} + \tfrac{1}{\sqrt2}\, \mathbb B_{\mathit\Sigma^+\to p\pi^0}^{} \,=\, (61.45\pm0.22)\times10^{-7} \,,
\nonumber \\
{\tt B}_{1,3}^{} & \,=\, \tfrac{2}{3}\, \mathbb B_{\mathit\Sigma^-\to n\pi^-}^{} + \tfrac{1}{3}\, \mathbb B_{\mathit\Sigma^+\to n\pi^+}^{} - \tfrac{\sqrt2}{3}\, \mathbb B_{\mathit\Sigma^+\to p\pi^0}^{}
\,=\, (0.26\pm0.15)\times10^{-7} \,,
\nonumber \\
{\tt B}_{3,3}^{} & \,=\, \sqrt{\tfrac{5}{18}} \Big(
\mathbb B_{\mathit\Sigma^-\to n\pi^-}^{} - \mathbb B_{\mathit\Sigma^+\to n\pi^+}^{} + \sqrt2~ \mathbb B_{\mathit\Sigma^+\to p\pi^0}^{} \Big) \,=\, (-2.67\pm0.19)\times10^{-7} \,. &
\end{align}
Evidently, $\mathbb A_{\mathit\Sigma^+\to n\pi^+}$ and $\mathbb B_{\mathit\Sigma^-\to n\pi^-}$ are merely a few percent in size of their counterparts in the other modes.
Moreover, the \,$\Delta I=1/2$\, component ${\tt B}_{1,3}$ is unexpectedly suppressed due to the observed smallness of $\mathbb B_{\mathit\Sigma^-\to n\pi^-}$.
Nevertheless, the \,$\Delta I=1/2$\, rule appears to hold in the light of
\begin{align}
\frac{{\tt A}_{3,3}}{\mathbb A_{\mathit\Sigma^-\to n\pi^-}} & \,=\, -0.045\pm0.004 \,, &
\frac{{\tt B}_{3,3}}{\mathbb B_{\mathit\Sigma^+\to n\pi^+}} & \,=\, -0.063\pm0.004 \,. &
\end{align}

\begin{table}[t] \centering
\caption{The latest data on $\alpha$, $\gamma$, and branching fraction $\mathscr B$ of \,$\mathit\Sigma\to N\pi$\, and the resulting $\mathbb A$ and $\mathbb B$, their phases having been ignored.
The $\alpha$ values are averages, including antiparticle modes if available.
The $\mathbb A$ and $\mathbb B$ numbers are in units of $10^{-7}$.\label{data}\vspace{5pt}}

\begin{tabular}{c||ccc|cc} \hline \small Decay mode & ~ $\alpha_{\mathit\Sigma\to N\pi}^{}\vphantom{\int_|^|}$ & ~ $\gamma_{\mathit\Sigma\to N\pi}^{}$ & $\mathscr B_{\mathit\Sigma\to N\pi}^{}$ $(\%)$ & $\mathbb A_{\mathit\Sigma\to N\pi}^{}$ & $\mathbb B_{\mathit\Sigma\to N\pi}^{}$ \\
\hline \raisebox{17pt}{} $\begin{array}{rl}
\mathit\Sigma^+ \to \hspace{-3pt} & n\pi^+       \raisebox{9pt}{} \\
\mathit\Sigma^+ \to\hspace{-3pt} & p\pi^0 \raisebox{9pt}{} \\
\mathit\Sigma^- \to \hspace{-3pt} & n\pi^-       \raisebox{9pt}{}
\end{array}$ & $\begin{array}{rl}
 0.0514 \,\pm \hspace{-4pt} & 0.0029 \raisebox{9pt}{}  \\
-0.9868 \,\pm \hspace{-4pt} & 0.0019 \raisebox{9pt}{}  \\
-0.0681  \,\pm \hspace{-4pt} & 0.0077 \raisebox{9pt}{}
\end{array}$ & ~ $\begin{array}{r}
\!\!\! -0.97 \raisebox{9pt}{} \\
 0.16 \raisebox{9pt}{} \\
 0.98 \raisebox{9pt}{}
\end{array}$ & ~ $\begin{array}{rl}
48.43 \,\pm \hspace{-4pt}  & 0.30  \raisebox{9pt}{} \\
51.47 \,\pm \hspace{-4pt} & 0.30  \raisebox{9pt}{} \\
99.848 \,\pm \hspace{-4pt} & 0.005 \raisebox{9pt}{}
\end{array}$ ~ & $\begin{array}{rl}
 0.11 \,\pm \hspace{-4pt} & 0.01 \raisebox{9pt}{} \\
-3.20 \,\pm \hspace{-4pt} & 0.02 \raisebox{9pt}{} \\
 4.27\,\pm \hspace{-4pt} & 0.02 \raisebox{9pt}{}
\end{array}$ ~ & $\begin{array}{rl}
42.17 \,\pm \hspace{-4pt} & 0.15 \raisebox{9pt}{} \\
27.26 \,\pm \hspace{-4pt} & 0.21 \raisebox{9pt}{} \\
-1.43 \,\pm \hspace{-4pt} & 0.16 \raisebox{9pt}{}
\end{array}$ \\ \hline
\end{tabular}
\end{table}

\subsection{\textit{CP}-odd observables\label{cpvo}}

The $CP$-violating observables of interest here are~\cite{BESIII:2021ypr,Brown:1983wd,Chau:1983ei,Donoghue:1985ww,Donoghue:1986hh}
\begin{align} \label{cpa}
\hat A_{\mathit\Sigma\to N\pi}^{} =~ & \frac{ \alpha_{\mathit\Sigma\to N\pi}^{} + \alpha_{\,\overline{\!\mathit\Sigma}\to\overline N\overline\pi} }{  \alpha_{\mathit\Sigma\to N\pi}^{} - \alpha_{\,\overline{\!\mathit\Sigma}\to\overline N\overline\pi} } \,, &  \widetilde{\textsc a}_{\mathit\Sigma\to N\pi}^{} =~ &  \frac{ \alpha_{\mathit\Sigma\to N\pi}^{}\, \Gamma_{\mathit\Sigma\to N\pi}^{} + \alpha_{\,\overline{\!\mathit\Sigma}\to\overline N\overline\pi}\, \Gamma_{\,\overline{\!\mathit\Sigma}\to\overline N\overline\pi} }{ \alpha_{\mathit\Sigma\to N\pi}^{}\, \Gamma_{\mathit\Sigma\to N\pi}^{} - \alpha_{\,\overline{\!\mathit\Sigma}\to\overline N\overline\pi}\, \Gamma_{\,\overline{\!\mathit\Sigma}\to\overline N\overline\pi} } \,,
\nonumber \\
\hat B_{\mathit\Sigma\to N\pi}^{} =~ & \frac{ \beta_{\mathit\Sigma\to N\pi}^{} + \beta_{\,\overline{\!\mathit\Sigma}\to\overline N\overline\pi} }{ \alpha_{\mathit\Sigma\to N\pi}^{} - \alpha_{\,\overline{\!\mathit\Sigma}\to\overline N\overline\pi} } \,, & \widetilde{\textsc b}_{\mathit\Sigma\to N\pi}^{} =~ &  \frac{ \beta_{\mathit\Sigma\to N\pi}^{}\, \Gamma_{\mathit\Sigma\to N\pi}^{} + \beta_{\,\overline{\!\mathit\Sigma}\to\overline N\overline\pi}\, \Gamma_{\,\overline{\!\mathit\Sigma}\to\overline N\overline\pi} }{ \alpha_{\mathit\Sigma\to N\pi}^{}\, \Gamma_{\mathit\Sigma\to N\pi}^{} - \alpha_{\,\overline{\!\mathit\Sigma}\to\overline N\overline\pi}\, \Gamma_{\,\overline{\!\mathit\Sigma}\to\overline N\overline\pi} } \,, ~~~
\nonumber \\
\Delta\phi _{\mathit\Sigma\to N\pi}^{} =~ & \frac{\phi_{\mathit\Sigma\to N\pi}^{} + \phi_{\,\overline{\!\mathit\Sigma}\to\overline N\overline\pi}}{2} \,, &
\hat\Delta_{\mathit\Sigma\to N\pi}^{} =~ & \frac{ \Gamma_{\mathit\Sigma\to N\pi}^{}
- \Gamma_{\,\overline{\!\mathit\Sigma}\to\overline N\overline\pi} }{ \Gamma_{\mathit\Sigma\to N\pi}^{}
+ \Gamma_{\,\overline{\!\mathit\Sigma}\to\overline N\overline\pi} } \,,
\end{align}
where ~$\overline{\!\mathit\Sigma}\to\,\overline{\!N}\overline\pi$\, is the antiparticle counterpart of \,$\mathit\Sigma\to N\pi$.\,
For \,$\mathit\Sigma^-\to n\pi^-$\, and the antihyperon mode \,$\overline{\!\mathit\Sigma}{}^+\to\overline n\pi^+$,\, both of which have final states with only one isospin value of \,$I_{\rm f}=3/2$,\, the $\mathbb A$ and $\mathbb B$ expressions can be seen to satisfy \,$|\mathbb A_{\mathit\Sigma^-\to n\pi^-}| =
|\mathbb A_{\,\overline{\!\mathit\Sigma}{}^+\to\overline n\pi^+}|$\,
and \,$|\mathbb B_{\mathit\Sigma^-\to n\pi^-}| =
|\mathbb B_{\,\overline{\!\mathit\Sigma}{}^+\to\overline n\pi^+}|$,\, as also pointed out in appendix~\ref{app:Sigma-}.
This implies that \,$\Gamma_{\mathit\Sigma^-\to n\pi^-}=\Gamma\raisebox{1pt}{$_{\,\overline{\!\mathit\Sigma}{}^+\to\overline n\pi^+}$}$\, and hence
\begin{align}
\widetilde{\textsc a}_{\mathit\Sigma^-\to n\pi^-} & \,=\, \hat A_{\mathit\Sigma^-\to n\pi^-} \,, & \widetilde{\textsc b}_{\mathit\Sigma^-\to n\pi^-} & \,=\, \hat B_{\mathit\Sigma^-\to n\pi^-} \,, & \hat\Delta_{\mathit\Sigma^-\to n\pi^-} & \,=\, 0 \,. ~~~
\end{align}
Furthermore, it is straightforward to arrive at
\begin{align} \label{ASm}
\hat A_{\mathit\Sigma^-\to n\pi^-}^{} & \,=\, \hat B_{\mathit\Sigma^-\to n\pi^-}^{} {\rm\,tan}\big(\delta_3^{\rm S}-\delta_3^{\rm P}\big) \,, &
\end{align}
where
\begin{align}
\hat B_{\mathit\Sigma^-\to n\pi^-}^{} & \,=\, \frac{ {\tt A}_{1,3}^{}\, {\tt B}_{1,3}^{}\, \hat{\tt S}_{13,13}^{\rm P,S} + \frac{\sqrt2}{\sqrt5} \Big( {\tt A}_{1,3}^{}\, {\tt B}_{3,3}^{}\, \hat{\tt S}_{33,13}^{\rm P,S} + {\tt A}_{3,3}^{}\, {\tt B}_{1,3}^{}\, \hat{\tt S}_{13,33}^{\rm P,S} \Big) + \frac{2}{5}\, {\tt A}_{3,3}^{}\, {\tt B}_{3,3}^{}\, \hat{\tt S}_{33,33}^{\rm\,P,S} }{ {\tt A}_{1,3}^{}\, {\tt B}_{1,3}^{}\, \hat{\tt C}_{13,13}^{\rm P,S} + \frac{\sqrt2}{\sqrt5} \Big( {\tt A}_{1,3}^{}\, {\tt B}_{3,3}^{}\, \hat{\tt C}_{33,13}^{\rm P,S} + {\tt A}_{3,3}^{}\, {\tt B}_{1,3}^{}\, \hat{\tt C}_{13,33}^{\rm P,S} \Big) + \frac{2}{5}\, {\tt A}_{3,3}^{}\, {\tt B}_{3,3}^{}\, \hat{\tt C}_{33,33}^{\rm\,P,S} } \,,
\end{align}
with \,$\hat{\tt C}_{x,z}^{\tt X,Z}\equiv{\rm cos}\big(\xi_x^{\tt X}-\xi_z^{\tt Z}\big)$\, and \,$\hat{\tt S}_{x,z}^{\tt X,Z}\equiv{\rm sin}\big(\xi_x^{\tt X}-\xi_z^{\tt Z}\big)$.\,

In the $\mathit\Sigma^+$ case, the asymmetries are more complicated, some of which are linked according to
\begin{align} \label{ABD}
\hat A_{\mathit\Sigma^+\to n\pi^+}^{} & \,=\, \frac{\widetilde{\textsc a}_{\mathit\Sigma^+\to n\pi^+} - \hat\Delta_{\mathit\Sigma^+\to n\pi^+}}{1 - \widetilde{\textsc a}_{\mathit\Sigma^+\to n\pi^+}\, \hat\Delta_{\mathit\Sigma^+\to n\pi^+}} \,, &
\hat B_{\mathit\Sigma^+\to n\pi^+}^{} & \,=\, \frac{\widetilde{\textsc b}_{\mathit\Sigma^+\to n\pi^+} - {\cal E}_{\mathit\Sigma^+\to n\pi^+}\, \hat\Delta_{\mathit\Sigma^+\to n\pi^+}}{1 - \widetilde{\textsc a}_{\mathit\Sigma^+\to n\pi^+}\, \hat\Delta_{\mathit\Sigma^+\to n\pi^+}} \,, ~~~
\nonumber \\
\hat A_{\mathit\Sigma^+\to p\pi^0}^{} & \,=\, \frac{\widetilde{\textsc a}_{\mathit\Sigma^+\to p\pi^0} -\hat\Delta_{\mathit\Sigma^+\to p\pi^0}}{1 - \widetilde{\textsc a}_{\mathit\Sigma^+\to p\pi^0}\, \hat\Delta_{\mathit\Sigma^+\to p\pi^0}} \,, &
\hat B_{\mathit\Sigma^+\to p\pi^0}^{} & \,=\, \frac{\widetilde{\textsc b}_{\mathit\Sigma^+\to p\pi^0} - {\cal E}_{\mathit\Sigma^+\to p\pi^0}\, \hat\Delta_{\mathit\Sigma^+\to p\pi^0}}{1 - \widetilde{\textsc a}_{\mathit\Sigma^+\to p\pi^0}\, \hat\Delta_{\mathit\Sigma^+\to p\pi^0}} \,, ~
\end{align}
where, with \,$\widetilde{\tt c}_{x,z}^{\mathscr{X,Z}} \equiv \cos\big(\delta_x^\mathscr X-\delta_z^\mathscr Z\big)$\, and \,$\widetilde{\tt s}_{x,z}^{\mathscr{X,Z}} \equiv \sin\big(\delta_x^\mathscr X-\delta_z^\mathscr Z\big)$,\,
\begin{align}
\widetilde{\textsc a}_{\mathit\Sigma^+\to n\pi^+}^{} & \,=\, \frac{-{\tt A_1^{}B_1^{}}\, \widetilde{\tt s}_{1,1}^{\rm P,S}\, \hat{\tt S}_{1,1}^{\rm P,S} - \tfrac{1}{2} {\tt A_1^{}B_3^{}}\, \widetilde{\tt s}_{3,1}^{\rm P,S}\, \hat{\tt S}_{3,1}^{\rm P,S} - \tfrac{1}{2} {\tt A_3^{}B_1^{}}\, \widetilde{\tt s}_{1,3}^{\rm P,S}\, \hat{\tt S}_{1,3}^{\rm P,S} - \tfrac{1}{4} {\tt A_3^{}B_3^{}}\, \widetilde{\tt s}_{3,3}^{\rm P,S}\, \hat{\tt S}_{3,3}^{\rm P,S}}{\mathscr D_{\mathit\Sigma^+\to n\pi^+}} \,, ~~~
\nonumber \\
\mathscr D_{\mathit\Sigma^+\to n\pi^+}^{} & \,=\, {\tt A_1^{}B_1^{}}\, \widetilde{\tt c}_{1,1}^{\rm P,S}\, \hat{\tt C}_{1,1}^{\rm P,S} + \tfrac{1}{2} {\tt A_1^{}B_3^{}}\, \widetilde{\tt c}_{3,1}^{\rm P,S}\, \hat{\tt C}_{3,1}^{\rm P,S} + \tfrac{1}{2} {\tt A_3^{}B_1^{}}\, \widetilde{\tt c}_{1,3}^{\rm P,S}\, \hat{\tt C}_{1,3}^{\rm P,S} + \tfrac{1}{4} {\tt A_3^{}B_3^{}}\, \widetilde{\tt c}_{3,3}^{\rm P,S}\, \hat{\tt C}_{3,3}^{\rm P,S} \,,
\nonumber \\
\hat\Delta_{\mathit\Sigma^+\to n\pi^+}^{} & \,=\, \frac{ -{\tt A}_1^{} {\tt A}_3^{}\, \widetilde{\tt s}_{1,3}^{\rm S,S}\, \hat{\tt S}_{1,3}^{\rm S,S} - {\tt B}_1^{} {\tt B}_3^{}\, \widetilde{\tt s}_{1,3}^{\rm P,P}\, \hat{\tt S}_{1,3}^{\rm P,P}\, \mathscr K_{\mathit\Sigma^+n\pi^+}^2 }{ {\tt A}_1^2 + \frac{1}{4}\, {\tt A}_3^2 + {\tt A}_1^{} {\tt A}_3^{}\,
\widetilde{\tt c}_{1,3}^{\rm S,S}\, \hat{\tt C}_{1,3}^{\rm S,S} + \Big( {\tt B}_1^2 + \frac{1}{4}\, {\tt B}_3^2 + {\tt B}_1^{} {\tt B}_3^{}\, \widetilde{\tt c}_{1,3}^{\rm P,P}\, \hat{\tt C}_{1,3}^{\rm P,P} \Big)
\mathscr K_{\mathit\Sigma^+n\pi^+}^2 } \,, ~~~ ~~
\nonumber \\
\widetilde{\textsc b}_{\mathit\Sigma^+\to n\pi^+}^{} & \,=\, \frac{{\tt A_1^{}B_1^{}}\, \widetilde{\tt c}_{1,1}^{\rm P,S}\, \hat{\tt S}_{1,1}^{\rm P,S} + \tfrac{1}{2} {\tt A_1^{}B_3^{}}\, \widetilde{\tt c}_{3,1}^{\rm P,S}\, \hat{\tt S}_{3,1}^{\rm P,S} + \tfrac{1}{2} {\tt A_3^{}B_1^{}}\, \widetilde{\tt c}_{1,3}^{\rm P,S}\, \hat{\tt S}_{1,3}^{\rm P,S} + \tfrac{1}{4} {\tt A_3^{}B_3^{}}\, \widetilde{\tt c}_{3,3}^{\rm P,S}\, \hat{\tt S}_{3,3}^{\rm P,S}}{\mathscr D_{\mathit\Sigma^+\to n\pi^+}} \,,
\nonumber \\
{\cal E}_{\mathit\Sigma^+\to n\pi^+}^{} & \,=\, \frac{{\tt A_1^{}B_1^{}}\, \widetilde{\tt s}_{1,1}^{\rm P,S}\, \hat{\tt C}_{1,1}^{\rm P,S} + \tfrac{1}{2} {\tt A_1^{}B_3^{}}\, \widetilde{\tt s}_{3,1}^{\rm P,S}\, \hat{\tt C}_{3,1}^{\rm P,S} + \tfrac{1}{2} {\tt A_3^{}B_1^{}}\, \widetilde{\tt s}_{1,3}^{\rm P,S}\, \hat{\tt C}_{1,3}^{\rm P,S} + \tfrac{1}{4} {\tt A_3^{}B_3^{}}\, \widetilde{\tt s}_{3,3}^{\rm P,S}\, \hat{\tt C}_{3,3}^{\rm P,S}}{\mathscr D_{\mathit\Sigma^+\to n\pi^+}} \,,
\end{align}
\begin{align}
\widetilde{\textsc a}_{\mathit\Sigma^+\to p\pi^0}^{} & \,=\, \frac{-{\tt A_1^{}B_1^{}}\, \widetilde{\tt s}_{1,1}^{\rm P,S}\, \hat{\tt S}_{1,1}^{\rm P,S} + {\tt A_1^{}B_3^{}}\, \widetilde{\tt s}_{3,1}^{\rm P,S}\, \hat{\tt S}_{3,1}^{\rm P,S} + {\tt A_3^{}B_1^{}}\, \widetilde{\tt s}_{1,3}^{\rm P,S}\, \hat{\tt S}_{1,3}^{\rm P,S} - {\tt A_3^{}B_3^{}}\, \widetilde{\tt s}_{3,3}^{\rm P,S}\, \hat{\tt S}_{3,3}^{\rm P,S}}{\mathscr D_{\mathit\Sigma^+\to p\pi^0}} \,, ~~~
\nonumber \\
\mathscr D_{\mathit\Sigma^+\to p\pi^0}^{} & \,=\, {\tt A_1^{}B_1^{}}\, \widetilde{\tt c}_{1,1}^{\rm P,S}\, \hat{\tt C}_{1,1}^{\rm P,S} - {\tt A_1^{}B_3^{}}\, \widetilde{\tt c}_{3,1}^{\rm P,S}\, \hat{\tt C}_{3,1}^{\rm P,S} - {\tt A_3^{}B_1^{}}\, \widetilde{\tt c}_{1,3}^{\rm P,S}\, \hat{\tt C}_{1,3}^{\rm P,S} + {\tt A_3^{}B_3^{}}\, \widetilde{\tt c}_{3,3}^{\rm P,S}\, \hat{\tt C}_{3,3}^{\rm P,S} \,,
\nonumber \\
\hat\Delta_{\mathit\Sigma^+\to p\pi^0}^{} & \,=\, \frac{ 2\, {\tt A}_1^{} {\tt A}_3^{}\, \widetilde{\tt s}_{1,3}^{\rm S,S}\, \hat{\tt S}_{1,3}^{\rm S,S} + 2\, {\tt B}_1^{} {\tt B}_3^{}\, \widetilde{\tt s}_{1,3}^{\rm P,P}\, \hat{\tt S}_{1,3}^{\rm P,P}\, \mathscr K_{\mathit\Sigma^+p\pi^0}^2 }{ {\tt A}_1^2 + {\tt A}_3^2 - 2\, {\tt A}_1^{} {\tt A}_3^{}\, \widetilde{\tt c}_{1,3}^{\rm S,S}\, \hat{\tt C}_{1,3}^{\rm S,S} + \Big( {\tt B}_1^2 + {\tt B}_3^2 - 2\, {\tt B}_1^{} {\tt B}_3^{}\, \widetilde{\tt c}_{1,3}^{\rm P,P}\, \hat{\tt C}_{1,3}^{\rm P,P} \Big) \mathscr K_{\mathit\Sigma^+p\pi^0}^2 }  \,,
\nonumber \\
\widetilde{\textsc b}_{\mathit\Sigma^+\to p\pi^0}^{} & \,=\, \frac{{\tt A_1^{}B_1^{}}\, \widetilde{\tt c}_{1,1}^{\rm P,S}\, \hat{\tt S}_{1,1}^{\rm P,S} - {\tt A_1^{}B_3^{}}\, \widetilde{\tt c}_{3,1}^{\rm P,S}\, \hat{\tt S}_{3,1}^{\rm P,S} - {\tt A_3^{}B_1^{}}\, \widetilde{\tt c}_{1,3}^{\rm P,S}\, \hat{\tt S}_{1,3}^{\rm P,S} + {\tt A_3^{}B_3^{}}\, \widetilde{\tt c}_{3,3}^{\rm P,S}\, \hat{\tt S}_{3,3}^{\rm P,S}}{\mathscr D_{\mathit\Sigma^+\to p\pi^0}} \,,
\nonumber \\
{\cal E}_{\mathit\Sigma^+\to p\pi^0}^{} & \,=\, \frac{{\tt A_1^{}B_1^{}}\, \widetilde{\tt s}_{1,1}^{\rm P,S}\, \hat{\tt C}_{1,1}^{\rm P,S} - {\tt A_1^{}B_3^{}}\, \widetilde{\tt s}_{3,1}^{\rm P,S}\, \hat{\tt C}_{3,1}^{\rm P,S} - {\tt A_3^{}B_1^{}}\, \widetilde{\tt s}_{1,3}^{\rm P,S}\, \hat{\tt C}_{1,3}^{\rm P,S} + {\tt A_3^{}B_3^{}}\, \widetilde{\tt s}_{3,3}^{\rm P,S}\, \hat{\tt C}_{3,3}^{\rm P,S}}{\mathscr D_{\mathit\Sigma^+\to p\pi^0}} \,.
\end{align}
Since $\widetilde{\textsc a}_{\mathit\Sigma^+\to N\pi}$ and $\hat\Delta_{\mathit\Sigma^+\to N\pi}$ are each already suppressed, from eq.\,(\ref{ABD}) we infer
\begin{align}
\hat A_{\mathit\Sigma^+\to N\pi} & \,\simeq\, \widetilde{\textsc a}_{\mathit\Sigma^+\to  N\pi} - \hat\Delta_{\mathit\Sigma^+\to  N\pi} \,, & \hat B_{\mathit\Sigma^+\to  N\pi} & \,\simeq\, \widetilde{\textsc b}_{\mathit\Sigma^+\to  N\pi} - {\cal E}_{\mathit\Sigma^+\to  N\pi}\, \hat\Delta_{\mathit\Sigma^+\to N\pi} \,.
\end{align}
Thus, in numerical work concerning the $\mathit\Sigma^+$ channels it suffices to deal with just $\hat A_{\mathit\Sigma^+\to N\pi}$, $\hat B_{\mathit\Sigma^+\to N\pi}$, and $\hat\Delta_{\mathit\Sigma^+\to N\pi}$, besides $\Delta\phi_{\mathit\Sigma^+\to N\pi}$.

Most of these $CP$-odd observables contain the strong phases, whose values are~\cite{Salone:2022lpt}
\begin{align} \label{deltas}
\delta_1^{\rm S} & = 9.98\pm0.23 \,, & \delta_3^{\rm S} & = -10.70\pm0.13 \,, & \delta_1^{\rm P} & = -0.04\pm0.33 \,, & \delta_3^{\rm P} & = -3.27\pm0.15 \,,
\end{align}
all in degrees, from \,$N\pi\to N\pi$\, data analyses~\cite{Hoferichter:2015hva}.
The empirical isospin components ${\tt A}_x$ and ${\tt B}_x$ listed in eq.\,(\ref{xAxB}) are needed as well for predicting $\hat A_{\mathit\Sigma\to N\pi}$, $\hat B_{\mathit\Sigma\to N\pi}$, $\Delta\phi_{\mathit\Sigma\to N\pi}$, and $\hat\Delta_{\mathit\Sigma\to N\pi}$, including in the evaluations of the weak phases $\xi_x^{\rm S}$ and $\xi_x^{\rm P}$, which additionally depend on the underlying $CP$-violating interactions within or beyond the SM.
In numerical work, we will employ the exact formulas written down above for these asymmetries, as the various terms therein can be of the same order of magnitude.

\section{Standard model predictions\label{smpredict}}

To address hyperon $CP$-violation quantitatively, we adopt a chiral-Lagrangian approach, where the lightest baryon and meson fields are organized into the matrices~\cite{Bijnens:1985kj}
\begin{align} \label{fields}
B & \,= \left(\!\begin{array}{ccc} \displaystyle \frac{\mathit\Lambda^{\vphantom{|^|}}}{\sqrt6} + \frac{\mathit\Sigma^0}{\sqrt2} & \mathit\Sigma^+ & p \\ \mathit\Sigma^- & \displaystyle \frac{\mathit\Lambda}{\sqrt6} - \frac{\mathit\Sigma^0}{\sqrt2} & n \\ \mathit\Xi^- & \mathit\Xi^0 & \displaystyle -\frac{\sqrt2}{\sqrt3_{\vphantom{|}}}\,\mathit\Lambda \end{array}\!\right) , &
\varphi & \,= \left(\!\begin{array}{ccc} \displaystyle \frac{\eta_8^{\vphantom{|}}}{\sqrt3} + \pi^0
& \sqrt2\,\pi^+ & \sqrt2\,K^+ \\ \sqrt2\, \pi^- & \displaystyle \frac{\eta_8^{}}{\sqrt3} - \pi^0 & \sqrt2\, K^0 \\ \sqrt2\, K^- &
\sqrt2~\overline{\!K}{}^0 & \displaystyle \frac{-2\eta_8^{}}{\sqrt3_{\vphantom{|}}}
\end{array}\!\right) , ~
\nonumber \\
\overline{\!B} & \,=\, B^\dagger \gamma_0^{} \,, & \hat\Sigma & \,=\,  \xi^2 \,=\, e^{i\varphi/f_\pi^{}} \,,
\end{align}
which transform under the chiral-symmetry group \,${\rm SU}(3)_L\times{\rm SU}(3)_R$\, as
\begin{align}
B & \,\to\, \hat UB\,\hat U^\dagger \,, &
\,\overline{\!B} & \,\to\, \hat U\,\overline{\!B}\,\hat U^\dagger \,, &
\hat\Sigma & \,\to\, \hat L\hat\Sigma\hat R^\dagger \,, &
\xi & \,\to\, \hat L\xi\hat U^\dagger \,=\, \hat U\xi\hat R^\dagger \,, ~~~
\end{align}
where \,$f_\pi^{}=92.07$ MeV \cite{ParticleDataGroup:2024cfk} denotes the pion decay constant, \,$\hat U\in{\rm SU}(3)$\, is defined implicitly by the $\xi$ equation, and \,$\hat X\in{\rm SU}(3)_X$,\, $X=L,R$.\,
These matrices enter the lowest-order strong chiral Lagrangian~\cite{Bijnens:1985kj}
\begin{equation} \label{Lstrong} \begin{array}[b]{rl}
{\cal L}_{\rm s}^{} \,\supset\, {\rm Tr} \Big[ \!\! & \overline{\!B} \gamma_\mu^{} i \widetilde\partial^\mu B + \,\overline{\!B} \gamma_\mu^{}\gamma_5^{} \big( \big\{{\cal A}^\mu, B\big\} {\cal D} + \big[{\cal A}^\mu, B\big] {\cal F} \big) + \,\overline{\!B} \big( b_D^{} \big\{M_+, B\big\} + b_F^{} \big[M_+,B\big] \big)
\\ & \!+\,\, \tfrac{1}{2} B_0^{} f_\pi^2\, M_+^{} \Big] \,, \end{array}
\end{equation}
where \,$\widetilde\partial^\mu B\equiv\partial^\mu B + \frac{1}{2} \big[ \xi\,\partial^\mu \xi^\dagger + \xi^\dagger \partial^\mu \xi, B \big]$,\, the parameters $\cal D$ and $\cal F$ $(b_D$ and $b_F)$ can be fixed from the data on semileptonic decays of the octet baryons (on their masses), \,${\cal A}^\kappa = \tfrac{i}{2} \big( \xi\,\partial^\kappa \xi^\dagger - \xi^\dagger \partial^\kappa \xi \big)$,\, and \,$M_+ = \xi^\dagger M_q^{} \xi^\dagger + \xi M_q^\dagger \xi$,\, with \,$M_q = {\rm diag}(m_u,m_d,m_s)$\, being the light-quark-mass matrix.

\begin{figure}[b] \centering
\includegraphics[trim=33mm 244mm 37mm 33mm,clip,width=0.98\textwidth]{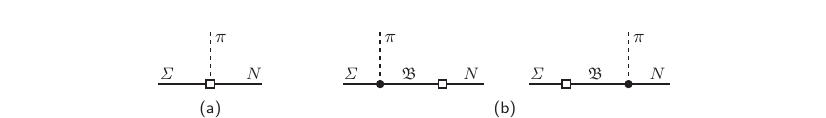}\vspace{-1ex}
\caption{Leading-order diagrams for the SM contributions to (a) {\tt S}- and (b) {\tt P}-wave \,$\mathit\Sigma\to N\pi$\, decay. Each hollow square symbolizes a coupling from ${\cal L}_{\scriptscriptstyle|\Delta S|=1}^{\tt SM}$ in eq.\,(\ref{Lw}).
In this and the next figure, each solid (dashed) line represents a spin-1/2 baryon (pseudoscalar meson), each thick dot a coupling from ${\cal L}_{\rm s}$ in~eq.\,(\ref{Lstrong}), and \,$\mathfrak B=N,\mathit\Lambda,$ or $\mathit\Sigma$.\,} \label{smdiagrams}
\end{figure}

In the SM the lowest-order weak chiral Lagrangian for \,$\Delta I=1/2$\, nonleptonic hyperon decays changing strangeness by \,$|\Delta S|$\,=\,1\, transforms as $(8,1)$ under \,${\rm SU}(3)_L\times{\rm SU}(3)_R$\, and has the form~\cite{Bijnens:1985kj}
\begin{align} \label{Lw}
{\cal L}_{\scriptscriptstyle|\Delta S|=1}^{\tt SM} & \,=\, {\rm Tr}\big( h_D^{}~ \overline{\!B} \big\{ \xi^\dagger\hat\kappa\xi,B \big\} + h_F^{}~ \overline{\!B} \big[ \xi^\dagger\hat\kappa\xi, B \big] \big) \,+\, \rm H.c. \,, &
\end{align}
which involves parameters $h_{D,F}$ and a 3$\times$3 matrix $\hat\kappa$ having elements \,$\hat\kappa_{kl}=\delta_{2k}\delta_{3l}$\,  projecting out \,$s\to d$\, transitions.
The ${\cal L}_{\scriptscriptstyle|\Delta S|=1}^{\tt SM}$ contributions to $\mathbb A$ and $\mathbb B$ at leading order are derived, respectively, from the contact diagram depicted in figure\,\ref{smdiagrams}(a) and from the baryon-pole diagrams in figure\,\ref{smdiagrams}(b).
The results are~\cite{Bijnens:1985kj}
\begin{align} \label{ABsm}
\mathbb A_{\mathit\Sigma^+\to n\pi^+}^{\tt SM} & \,=\, 0 \,, ~~~ ~~~~ ~~~ \mathbb A_{\mathit\Sigma^+\to p\pi^0}^{\tt SM} \,=\, \frac{h_D-h_F}{2 f\!_\pi^{}} \,=\, \frac{-\mathbb A_{\mathit\Sigma^-\to n\pi^-}^{\tt SM}}{\sqrt2}\,, &
\nonumber \\ \raisebox{4ex}{}
\mathbb B_{\mathit\Sigma^+\to n\pi^+}^{\tt SM} & \,=\, \frac{m_\mathit\Sigma+m_N}{3\sqrt2\, f\!_\pi^{}} \bigg( \frac{h_D+3h_F}{m_N-m_\mathit\Lambda} + 3\,\frac{h_D-h_F}{m_N-m_\mathit\Sigma} \bigg) \cal D \,,
\nonumber \\
\mathbb B_{\mathit\Sigma^+\to p\pi^0}^{\tt SM} & \,=\, \frac{m_\mathit\Sigma+m_N}{2 f\!_\pi^{}} \bigg( \frac{h_D-h_F}{m_N-m_\mathit\Sigma} \bigg) ({\cal D-F}) \,, &
\nonumber \\
\mathbb B_{\mathit\Sigma^-\to n\pi^-}^{\tt SM} & \,=\, \mathbb B_{\mathit\Sigma^+\to n\pi^+}^{\tt SM} - \sqrt2~ \mathbb B_{\mathit\Sigma^+\to p\pi^0}^{\tt SM} \,,
\end{align}
where $m_N$ and $m_\mathit\Sigma$ are isospin-averaged masses of the nucleons and $\mathit\Sigma^{+,0,-}$, respectively.
From these, it is straightforward to deduce $\dot{\tt A}_{1,3}^{\tt SM}$, $\dot{\tt A}_1^{\tt SM}$, $\dot{\tt B}_{1,3}^{\tt SM}$, and $\dot{\tt B}_1^{\tt SM}$ in analogy to their empirical counterparts in eq.\,(\ref{xAxB}) and \,$\dot{\tt A}_{3,3}^{\tt SM} = \dot{\tt B}_{3,3}^{\tt SM}=0$,\, as ${\cal L}_{\scriptscriptstyle|\Delta S|=1}^{\tt SM}$ alters isospin solely by \,$\Delta I=1/2$.\,

These amplitudes originate from the quark-level \,$|\Delta S|=1$\, effective Hamiltonian~\cite{Buchalla:1995vs}
\begin{align} \label{Heffsm}
{\cal H}_{\rm eff}^{\tt SM} & =\, \frac{G_{\rm F}}{\sqrt2}~
\raisebox{3pt}{\footnotesize$\displaystyle\sum_{j=1}^{10}$} \Big( V_{ud}^*V_{us}^{}\, z_j^{}
- V_{td}^*V_{ts}^{}\, y_j^{} \Big) Q_j^{} \,+\, \rm H.c. \,, &
\end{align}
where \,$G_{\rm F}=1.1663788\times10^{-5}\rm\,GeV^{-2}$ \cite{ParticleDataGroup:2024cfk} is the Fermi coupling constant, $V_{kl}$ represent the elements of the Cabibbo-Kobayashi-Maskawa matrix, $Q_j^{}$ are four-quark operators whose expressions are written down in ref.\,\cite{Buchalla:1995vs}, and $z_j^{}$ and $y_j^{}$ stand for their Wilson coefficients and are real numbers.
The sources of $CP$ violation in ${\cal H}_{\rm eff}^{\tt SM}$ are then located in its $y_j^{}$ terms, primarily the one with $y_6^{}$ because it has the largest magnitude among $y_{1,\cdots,10}^{}$ and the baryonic matrix elements of the associated  QCD penguin operator $Q_6^{}=-8\big( \overline{d_L^{}}u_R^{}\, \overline{u_R^{}}s_L^{} + \overline{d_L^{}}d_R^{}\, \overline{d_R^{}}s_L^{} + \overline{d_L^{}}s_R^{}\, \overline{s_R^{}}s_L^{} \big)$\,  are bigger than those of $Q_{j\neq6}$.

To determine the $CP$-violating weak phases, which are much less than unity, following the usual practice~\cite{Donoghue:1986hh,He:1995na,Tandean:2002vy,Tandean:2003fr} we apply the approximations  \,$\xi_x^{\rm S}=\big({\rm Im}\,\dot{\tt A}_x^{\rm th}\big)/{\tt A}_x^{\rm exp}$\, and \,$\xi_x^{\rm P}=\big({\rm Im}\,\dot{\tt B}_x^{\rm th}\big)/{\tt B}_x^{\rm exp}$,\, where $\dot{\tt A}_x^{\rm th}$ and
$\dot{\tt B}_x^{\rm th}$ are the theoretical counterparts of the experimental isospin components ${\tt A}_x^{\rm exp}$ and
${\tt B}_x^{\rm exp}$, respectively, quoted in eq.\,(\ref{xAxB}).
Hence the weak phases in the SM are obtained using \,Im\,$\dot{\tt A}_x^{\tt SM}$ and \,Im\,$\dot{\tt B}_x^{\tt SM}$, which are linked to \,Im\,$h_{D,F}$ as indicated above.

In keeping with ref.\,\cite{Tandean:2002vy}, we work out the effects of $Q_6$ on $h_{D,F}$ from the factorization contributions, treating $Q_6^{}$ as comprising a sum of the products of two (pseudo)scalar quark bilinears, and from the nonfactorization contributions estimated in the MIT bag model~\cite{Donoghue:2022wrw}.
The results for the former are
\begin{align} \label{factoriz}
{\rm Im}\, h_D^{\rm fac} & \,=\, 4\sqrt2\, \eta_{\tt W}^{}\lambda_{\tt W}^5A_{\tt W}^2\, b_D^{} B_0^{} f_\pi^2 G_{\rm F}^{} y_6^{} \,, & {\rm Im}\, h_F^{\rm fac} & \,=\, 4\sqrt2\, \eta_{\tt W}^{}\lambda_{\tt W}^5A_{\tt W}^2\, b_F^{} B_0^{} f_\pi^2 G_{\rm F}^{} y_6^{} \,, &
\end{align}
and for the latter
\begin{align} \label{nonfact}
{\rm Im}\, h_D^{\rm nonfac} & \,=\, \tfrac{1}{\sqrt2} (3a'-5b') \eta_{\tt W}^{}\lambda_{\tt W}^5A_{\tt W}^2\, G_{\rm F}^{} y_6^{} \,, & {\rm Im}\, h_F^{\rm nonfac} & \,=\, \tfrac{1}{\sqrt2} \Big( a' + \tfrac{11}{3} b' \Big) \eta_{\tt W}^{}\lambda_{\tt W}^5A_{\tt W}^2\, G_{\rm F}^{} y_6^{} \,, &
\end{align}
where \,$\lambda_{\tt W} = 0.22501$,\, $A_{\tt W} = 0.826$,\, and \,$\eta_{\tt W}^{} = 0.361$\, from ref.\,\cite{ParticleDataGroup:2024cfk} are the Wolfenstein parameters~\cite{Wolfenstein:1983yz}, $b_D^{}=0.226$\, and \,$b_F^{}=-0.811$\, from fitting to the octet baryons' observed masses, \,$y_6^{}=-0.113$\,  from ref.\,\cite{Buchalla:1995vs},\footnote{This $y_6^{}$ number is the LO value at \,$\mu=1$ GeV\, for
\,$\Lambda_{\scriptscriptstyle\overline{\rm MS}}^{(4)}=325\,\rm MeV$\, itemized in Table XVIII of ref.\,\cite{Buchalla:1995vs}.\medskip} \,$a' = 0.00140\rm\,GeV^3$\, and \,$b' = 0.00064\rm\,GeV^3$\, are bag-model parameters from ref.\,\cite{Tandean:2002vy}, and $B_0^{}=m_K^2/(\hat m\!+\!m_s)$\,  with \,$\hat m=(m_u\!+\!m_d)/2$\, and \,$(m_u,m_d,m_s)=(2.85,6.20,123)$\,MeV\, at the renormalization scale \,$\mu=1$ GeV.\,
Accordingly, \,${\rm Im}\,h_{D(F)}^{\rm fac}\simeq 29\,(-28)\,{\rm Im}\,h_{D(F)}^{\rm nonfac}$.\,
Additional relevant constants are \,${\cal D}=0.81$\, and \,${\cal F}=0.47$\, inferred at lowest order from the data on semileptonic octet-baryon decays~\cite{ParticleDataGroup:2024cfk}.

\begin{figure}[h] \bigskip \centering
\includegraphics[height=3in]{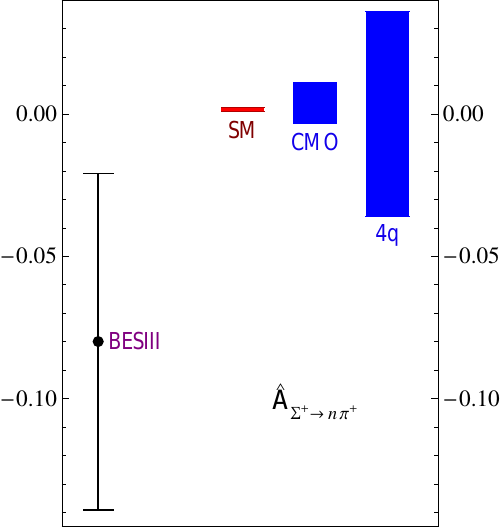} ~ ~ ~ \includegraphics[height=3in]{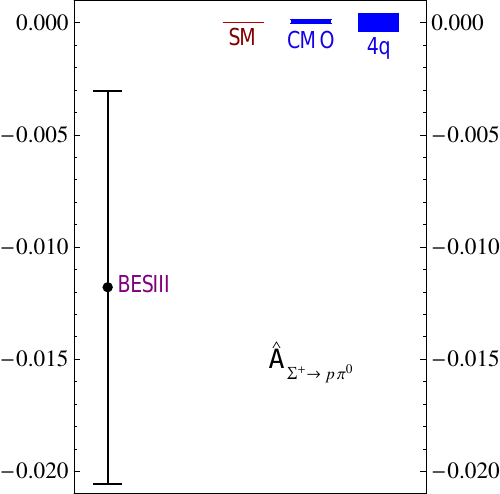}
\caption{The predicted $\hat A_{\mathit\Sigma^+\to n\pi^+}$ and $\hat A_{\mathit\Sigma^+\to p\pi^0}$ in the SM (red) and in the new-physics scenarios (blue) dealt with in sections~\ref{cmo} (\textsf{\small CMO}) and~\ref{dmmodel} (\textsf{\small 4q}), compared to the corresponding 1$\sigma$ (black) intervals of the BESIII findings cited in eq.\,(\ref{ASigma-data}), the statistical and systematical errors having been summed in quadrature.} \label{A-exp-th}
\end{figure}

With the central values of the input parameters, we then have, in units of  $\eta_{\tt W}^{}\lambda_{\tt W}^5A_{\tt W}^2$,
\begin{align} \label{xi-sm}
\xi_1^{\rm S,\tt SM} & \,=\, 2.2 \,, & \xi_{13}^{\rm S,\tt SM} & \,=\, 2.2 \,, & \xi_1^{\rm P,\tt SM} & \,=\, 0.22 \,, &
\xi_{13}^{\rm P,\tt SM} & \,=\, -103 \,, ~~~ ~~~~
\end{align}
and \,$\xi_{33}^{\rm S,\tt SM} = \xi_{33}^{\rm P,\tt SM}=0$.\,
As discussed in ref.\,\cite{Tandean:2002vy}, these results have sizable relative uncertainties, which we take here to be of order unity.\footnote{This is also in line with what was learned from one-loop computations of the leading nonanalytic corrections to the lowest-order amplitudes for nonleptonic hyperon decays in chiral perturbation theory~\cite{Bijnens:1985kj,Jenkins:1991bt,Springer:1999sv,AbdEl-Hady:1999llb}: specifically, the chiral logarithmic correction is of order one compared to the leading tree-level contribution, and the neglected ${\cal O}(m_s)$ correction is similar in size to the calculated ${\cal O}(m_s\ln m_s)$ correction.}
To incorporate this, along with \,$\eta_{\tt W}^{}\lambda_{\tt W}^5A_{\tt W}^2=1.42\times10^{-4}$,\, we collect $10^5$ sets of the weak-phases in eq.\,(\ref{xi-sm}) which are randomly generated from a uniform distribution within their respective assigned ranges.
After including them, as well as the central values of the components in eq.\,(\ref{xAxB}) and of the strong phases in eq.\,(\ref{deltas}), into the $CP$ asymmetries, we arrive at the latter's averages and 1$\sigma$ errors:
\begin{align} \label{cpa-sm}
\hat A_{\mathit\Sigma^+\to n\pi^+}^{\tt SM} & \,=\, (1.6\pm0.7) \times 10^{-3} \,, &
\hat B_{\mathit\Sigma^+\to n\pi^+}^{\tt SM} & \,=\, (-0.1\pm3.5) \times 10^{-3} \,,
\nonumber \\
\hat A_{\mathit\Sigma^+\to p\pi^0}^{\tt SM} & \,=\, (1.6\pm1.0) \times 10^{-5} \,, &
\hat B_{\mathit\Sigma^+\to p\pi^0}^{\tt SM} & \,=\, (-2.0\pm1.4) \times 10^{-4} \,, ~~~~~
\nonumber \\
\hat A_{\mathit\Sigma^-\to n\pi^-}^{\tt SM} & \,=\, (-3.0\pm2.0) \times 10^{-4} \,,  &
\hat B_{\mathit\Sigma^-\to n\pi^-}^{\tt SM} & \,=\, (2.3\pm1.5) \times 10^{-3} \,,
\nonumber \\
\Delta\phi_{\mathit\Sigma^+\to n\pi^+}^{\tt SM} & \,=\, (0.0\pm1.7) \times 10^{-4} \,,  &
\hat\Delta_{\mathit\Sigma^+\to n\pi^+}^{\tt SM} & \,=\, (-0.2\pm2.3) \times 10^{-5} \,,
\nonumber \\
\Delta\phi_{\mathit\Sigma^+\to p\pi^0}^{\tt SM} & \,=\, (1.1\pm0.7) \times 10^{-3} \,,  &
\hat\Delta_{\mathit\Sigma^+\to p\pi^0}^{\tt SM} & \,=\, (0.2\pm2.3) \times 10^{-5} \,,
\nonumber \\
\Delta\phi_{\mathit\Sigma^-\to n\pi^-}^{\tt SM} & \,=\, (-1.6\pm1.0) \times 10^{-4} \,.
\end{align}  
It is interesting to notice that $\hat A_{\mathit\Sigma^+\to n\pi^+}^{\tt SM}$ and $\hat A_{\mathit\Sigma^+\to p\pi^0}^{\tt SM}$  are close to the ranges of the corresponding data listed in eq.\,(\ref{ASigma-data}) and agree with them at the 2$\sigma$ level.
This is illustrated in figure~\ref{A-exp-th}, where the red patches exhibit the SM predictions and the black error-bars span the 1$\sigma$ intervals of the BESIII measurements.
Furthermore, the $\hat\Delta$ numbers in eq.\,(\ref{cpa-sm}) are consistent with the general expectation~\cite{Brown:1983wd} that \,$\hat\Delta_{\mathit\Sigma^+\to n\pi^+} \simeq -\hat\Delta_{\mathit\Sigma^+\to p\pi^0}$,\, which is based on the relations \,$\Gamma_{\mathit\Sigma^+\to n\pi^+} \simeq \Gamma_{\mathit\Sigma^+\to p\pi^0} \simeq \Gamma_{\mathit\Sigma^+}/2$\, from experiment~\cite{ParticleDataGroup:2024cfk} and \,$\Gamma_{\mathit\Sigma^+}^{}=\Gamma_{\overline{\!\mathit\Sigma}{}^-}$\, from the $CPT$ theorem.

\section{New physics scenarios\label{np1}}

The impact of possible new-physics on the hyperon $CP$-odd observables has been theoretically explored before to varying extents.
Especially, it was pointed out early on~\cite{Donoghue:1985ww,Donoghue:1986hh} that these could be enlarged in multi-Higgs models, via the chromomagnetic-penguin operators (CMOs).
A~model-independent investigation~\cite{He:1995na} later confirmed that they could indeed create some of the strongest effects consistent with kaon constraints and moreover demonstrated that scalar four-quark operators could likewise give rise to highly amplified contributions.
The consequences of the CMOs induced by new physics have since been addressed more extensively~\cite{Deshpande:1994vp,Chang:1994wk,He:1999bv,Chen:2001cv,Tandean:2003fr,Tandean:2004mv}, the majority of the studies concentrating on \,$\mathit\Lambda\to N\pi$\, and \,$\mathit\Xi\to\mathit\Lambda\pi$,\, motivated by previous experimental searches.
In this section we first revisit this kind of scenario to examine how it could bring about substantial $CP$-violation in \,$\mathit\Sigma\to N\pi$.\,
Subsequently we perform an analogous analysis in the context of a particular model which provides certain scalar four-quark operators that have recently been shown to cause significant $CP$-asymmetries in the $\mathit\Lambda$ and $\mathit\Xi$ instances.

\subsection{Enhanced chromomagnetic-penguin interactions\label{cmo}}

Model-independently, the low-energy effective Lagrangian for the CMOs is expressible as
\begin{equation}  \label{dsG}
\mathfrak L_{dsG}^{} \,=\, \frac{-g_{\rm s}^{}}{16\pi^2}\, \overline d \big( \textsl{\texttt C}_g P_R^{} + \,\widetilde{\!\textsl{\texttt C}}_g P_L^{} \big) \lambda_a^{}\textsl{\texttt G}_a^{\nu\tau} \sigma_{\nu\tau}^{}\, s \,+\, \rm H.c. \,, ~~~
\end{equation}
where  $g_{\rm s}^{}$ denotes the strong coupling, the Wilson coefficients $\textsl{\texttt C}_g$ and $\,\widetilde{\!\textsl{\texttt C}}_g$ are in general complex and unrelated to each other, $\lambda_a^{}\textsl{\texttt G}_a^{\nu\tau}$ is a Gell-Mann matrix acting on color space times the gluon field-strength tensor, summation over \,$a=1,2,...,8$\, being implicit, and \,$\sigma_{\nu\tau}=(i/2)\big[\gamma_\nu^{},\gamma_\tau^{}\big]$.
The leading-order chiral realization of $\mathfrak L_{dsG}$ translates into the effective Lagrangian~\cite{Tandean:2003fr}
\begin{align} \label{Lgnew}
\mathfrak L_\chi^{\tt CMO}  & \supset\, {\rm Tr} \Big[ \,\overline{\!B} \Big\{ \beta_D^{}\, \xi^\dagger \hat\kappa \xi^\dagger + \widetilde\beta_D^{}\, \xi \hat\kappa \xi, B \Big\} + \,\overline{\!B} \Big[ \beta_F^{}\, \xi^\dagger \hat\kappa \xi^\dagger + \widetilde\beta_F^{}\, \xi \hat\kappa \xi, B \Big]
+ B_0^{} f_\pi^2 \hat\kappa \Big( \beta_\varphi^{} \hat\Sigma^\dagger + \widetilde\beta_\varphi^{} \hat\Sigma \Big) \Big] ~~
\nonumber \\ & ~~~~ +\, \rm H.c. \,,
\end{align}
where the parameters $\beta_{D,F,\varphi}$ and $\widetilde\beta_{D,F,\varphi}$
are proportional to $\textsl{\texttt C}_g$ and $\,\widetilde{\!\textsl{\texttt C}}_g$, respectively.
It is clear from eq.\,(\ref{dsG}) that these interactions change isospin by \,$\Delta I=1/2$.\,

Anticipating how the quark operators in $\mathfrak L_{dsG}$ impact different hyperon and kaon observables, we  rearrange it as \,$\mathfrak L_{dsG}^{} = -\textsl{\texttt C}_g^+\textsl{\texttt Q}_g^+ -\textsl{\texttt C}_g^-\textsl{\texttt Q}_g^- + \rm H.c.$,\, where
\begin{align}
\textsl{\texttt C}_g^\pm & \,=\, \textsl{\texttt C}_g^{}\pm\,\widetilde{\!\textsl{\texttt C}}_g^{} \,, & \textsl{\texttt Q}_g^+ & \,=\, \frac{g_{\rm s}^{}}{32\pi^2}\, \overline d\, \lambda_a^{} \textsl{\texttt G}_a^{\nu\tau} \sigma_{\nu\tau}^{} s \,, & \textsl{\texttt Q}_g^- & \,=\, \frac{g_{\rm s}^{}}{32\pi^2}\, \overline d\, \lambda_a^{} \textsl{\texttt G}_a^{\nu\tau} \sigma_{\nu\tau}^{} \gamma_5^{} s \,. ~
\end{align}
We see that $\textsl{\texttt Q}_g^+$ $\big(\textsl{\texttt Q}_g^-\big)$ is even (odd) under parity and under a $CPS$ transformation, the latter being ordinary $CP$ followed by switching the $d$ and $s$ quarks.

\begin{figure}[b] \centering
\includegraphics[trim=39mm 244mm 38mm 37mm,clip,width=0.95\textwidth]{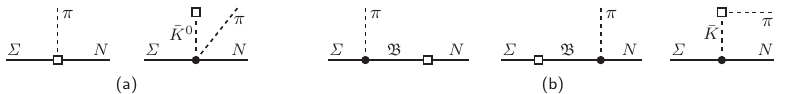}\vspace{-1ex}
\caption{Leading-order diagrams for the new contributions to (a) {\tt S}- and (b) {\tt P}-wave \,$\mathit\Sigma\to N\pi$\, decay. Each hollow square symbolizes a coupling from $\mathfrak L_\chi^{\tt CMO}$ in eq.\,(\ref{Lgnew}) or ${\cal L}_\chi^{\rm new}$ in eq.\,(\ref{Lchi}).} \label{npdiagrams}
\end{figure}

From $\mathfrak L_\chi^{\tt CMO}$, we can draw the contact and tadpole diagrams in figure\,\,\ref{npdiagrams}(a) and the baryon- and kaon-pole diagrams in figure\,\,\ref{npdiagrams}(b) representing its contributions to, respectively, the {\tt S}- and {\tt P}-wave amplitudes for \,$\mathit\Sigma\to N\pi$.\,
Thus, we have
\begin{align} \label{AB-cmo}
\mathbb A_{\mathit\Sigma^+\to n\pi^+}^{\tt CMO} & \,=\, 0 \,, ~~~ ~~~~ \mathbb A_{\mathit\Sigma^+\to p\pi^0}^{\tt CMO} \,=\,  \frac{\beta_D^--\beta_F^-}{2 f_\pi^{}} + \frac{\beta_\varphi^-}{2f_\pi^{}} \bigg(\frac{m_\mathit\Sigma^{}-m_N^{}}{\hat m-m_s^{}}\bigg) \,=\, \frac{-\mathbb A_{\mathit\Sigma^-\to n\pi^-}^{\tt CMO}}{\sqrt2} \,,
\nonumber \\
\mathbb B_{\mathit\Sigma^+\to n\pi^+}^{\tt CMO} & \,=\, \frac{m_N^{}+m_\mathit\Sigma^{}}{3\sqrt2\,f_\pi^{}} \Bigg( \frac{\beta_D^++3\beta_F^+}{m_N^{}-m_\mathit\Lambda^{}} + 3\,\frac{\beta_D^+-\beta_F^+}{m_N^{}-m_\mathit\Sigma^{}} \Bigg) \cal D \,,
\nonumber \\
\mathbb B_{\mathit\Sigma^+\to p\pi^0}^{\tt CMO} & \,=\,  \frac{m_N^{}+m_\mathit\Sigma^{}}{2f_\pi^{}} \Bigg( \frac{\beta_D^+-\beta_F^+}{m_N^{}-m_\mathit\Sigma^{}} + \frac{\beta_\varphi^+}{m_s^{}-\hat m} \Bigg)  (\cal D-F) \,,
\nonumber \\
\mathbb B_{\mathit\Sigma^-\to n\pi^-}^{\tt CMO} & \,=\, \mathbb B_{\mathit\Sigma^+\to n\pi^+}^{\tt CMO} - \sqrt2\, \mathbb B_{\mathit\Sigma^+\to p\pi^0}^{\tt CMO} \,,
\end{align}
where \,$\beta_X^\pm \equiv \beta_X^{} \pm \widetilde\beta_X^{}$\, and we have invoked the relations \,$m_N^{}-m_\mathit\Sigma^{}=2(b_D-b_F)\big(\hat m-m_s^{}\big)$\, and \,$m_\pi^2-m_K^2=B_0^{}\big(\hat m-m_s^{}\big)$ from eq.\,(\ref{Lstrong}), with $m_\pi$ and $m_K$ being the isospin-averaged masses of $\pi^{+,0,-}$ and $K^{+,0}$, respectively.

We note that $\mathbb A_{\mathit\Sigma\to N\pi}^{\tt CMO}$ and $\mathbb B_{\mathit\Sigma\to N\pi}^{\tt CMO}$ in eq.\,(\ref{AB-cmo}) would all become zero upon setting  $\beta_{D,F}^\pm = b_{D,F}^{}\,\textsc k^\pm$\, and \,$\beta_\varphi^\pm = \textsc k^\pm/2$,\, with $\textsc k^\pm$  being constants, and using the mass formulas mentioned in the preceding paragraph plus \,$m_N^{}-m_\Lambda^{} = 2\big(b_D^{}/3+b_F^{}\big) \big(m_s^{}-\hat m\big)$\,  from eq.\,(\ref{Lstrong}).
This complies with the requirement deduced from the Feinberg-Kabir-Weinberg
theorem~\cite{Feinberg:1959ui} that the operators \,$\overline d \big(1\pm\gamma_5^{}\big)s$\,
cannot contribute to physical amplitudes~\cite{Donoghue:1985ae,Donoghue:1986nv} and therefore serves
as a check for eq.\,(\ref{AB-cmo}).

From $\mathbb A_{\mathit\Sigma\to N\pi}^{\tt CMO}$ and $\mathbb B_{\mathit\Sigma\to N\pi}^{\tt CMO}$, we can derive the isospin components $\dot{\tt A}_{1,3}^{\tt CMO}$, $\dot{\tt A}_1^{\tt CMO}$, $\dot{\tt B}_{1,3}^{\tt CMO}$, and $\dot{\tt B}_1^{\tt CMO}$  in like manner to their empirical counterparts in eq.\,(\ref{xAxB}).
Then, employing the central values of the input parameters, along with
\,$\beta_D^\pm = -(3/7) \beta_F^\pm = 0.0011\, \textsl{\texttt C}_g^\pm\rm\,GeV^2$\, and \,$\beta_\varphi^\pm = -0.0037\,\textsl{\texttt C}_g^\pm\rm\,GeV^2$,\, updated from the estimates in ref.\,\cite{Tandean:2003fr}
based on the bag-model results of refs.\,\cite{Donoghue:1980rq,Donoghue:1976nb}, we find the CMO contributions to the weak phases to be
\begin{align} \label{xi-cmo}
\xi_1^{\rm S,\tt CMO} & \,=\, -2.1\times10^5 {\rm~GeV~Im}\,\textsl{\texttt C}_g^- \,, & \xi_{13}^{\rm S,\tt CMO} & \,=\, -2.0\times10^5 {\rm~GeV~Im}\,\textsl{\texttt C}_g^- \,, ~~~ ~~
\nonumber \\
\xi_1^{\rm P,\tt CMO} & \,=\, -2.5\times10^4 {\rm~GeV~Im}\,\textsl{\texttt C}_g^+ \,, & \xi_{13}^{\rm P,\tt CMO} & \,=\, 8.8\times10^6 {\rm~GeV~Im}\,\textsl{\texttt C}_g^+ \,,
\end{align}
and \,$\xi_{33}^{\rm S,\tt CMO}=\xi_{33}^{\rm P,\tt CMO}=0$.\,
These results are expected to be only accurate up to relative uncertainties of order unity, as was the case with the SM weak phases in section~\ref{smpredict}.

Since $\mathfrak L_{dsG}$ also affects the kaon decays \,$K\to\pi\pi$\, and neutral-kaon mixing, constraints from their measurements need to be taken into account.
With regard to the latter, the indirect $CP$-violation parameter $\varepsilon$ receives a contribution via long-distance effects~\cite{Donoghue:1985ae,He:1999bv}, mediated by mesons such as $\pi^0$, $\eta$, and $\eta'$.
Numerically, it can be written as~\cite{Tandean:2003fr}
\begin{equation}
\varepsilon_{\tt CMO}^{} \,=\,-2.3\times10^5\, \kappa {\rm~GeV~Im}\, \textsl{\texttt C}_g^+ \,, ~~~
\end{equation}
where $\kappa$ has a range given by \,$0.2<|\kappa|<1$\, and quantifies the impact of the different meson mediators~\cite{He:1999bv}.
The pertinent data and SM expectation are~\cite{ParticleDataGroup:2024cfk} \,$|\varepsilon_{\rm exp}|=(2.228\pm0.011)\times10^{-3}$ and~\cite{Gorbahn:2024qpe} \,$|\varepsilon_{\textsc{sm}}|=(2.171\pm0.181)\times10^{-3}$,\, respectively, the perturbative, nonperturbative, and parametric errors of $|\varepsilon_{\textsc{sm}}|$ having been combined in quadrature.
From the 2$\sigma$ range of \,$|\varepsilon_{\rm exp}|-|\varepsilon_{\textsc{sm}}|$,\, we may then demand \,$-3.1 < 10^4\, \varepsilon_{\tt CMO}^{} < 4.2$,\, which implies
\begin{equation} \label{ImCg+bound}
\big|{\rm Im}\,\textsl{\texttt C}_g^+\big| \,<\, 9.1\times10^{-9} \rm\,GeV^{-1} \,. ~~~
\end{equation}

Although the effect of $\mathfrak L_\chi^{\tt CMO}$ on \,$K\to\pi\pi$\, is known to vanish~\cite{Donoghue:1985ae}, nonzero contributions arise from the chiral realization of the $\textsl{\texttt Q}_g^-$ portion of $\mathfrak L_{dsG}$ at next-to-leading order, given by the correspondence~\cite{Bertolini:1994qk}
\begin{equation}
\textsl{\texttt Q}_g^- \,\Leftrightarrow\,
\frac{11 B_0^{}\, \textit{\texttt B}_{\tt CMO}^{}\, f_{\!\pi}^2}{128\pi^2} \big( \hat\Sigma^\dagger\,
\partial_\nu^{}\hat\Sigma\,\partial^\nu\hat\Sigma^\dagger - \partial_\nu^{}\hat\Sigma\,
\partial^\nu\hat\Sigma^\dagger\, \hat\Sigma \big)_{32} \,, ~~~
\end{equation}
with the $\textit{\texttt B}$ factor \,$\textit{\texttt B}_{\tt CMO}^{} = 0.273\pm0.069$\, from a lattice-QCD calculation~\cite{Constantinou:2017sgv}.
This leads to the amplitudes
\begin{equation}
-i{\cal M}_{K^0\to\pi^+\pi^-}^{\tt CMO} \,=\, -i{\cal M}_{K^0\to\pi^0\pi^0}^{\tt CMO}
\,=\, \frac{-11 B_0^{}\, \textit{\texttt B}_{\tt CMO}^{}\, \textsl{\texttt C}_g^{-*}\, m_\pi^2}
{32\sqrt2\,\pi^2 f\!_\pi^{}} \,=\, A_0^{\tt CMO} \,, ~~~
\end{equation}
where $A_0^{\tt CMO}$ stands for the CMO contribution to the \,$\Delta I=1/2$\, component $A_0$.
Ignoring the uncertainties of the various input parameters except $\textit{\texttt B}_{\tt CMO}$, we then obtain
\begin{align}
\frac{\varepsilon_{\tt CMO}'}{\varepsilon}\, & =\, \frac{-\omega~{\rm Im}A_0^{\tt CMO}}
{\sqrt2\,|\varepsilon_{\rm exp}|\, {\rm Re}A_0^{\rm exp}}
\,=\, \frac{-11 \omega B_0^{}\,  \textit{\texttt B}_{\tt CMO}^{}\, m_\pi^2\, {\rm Im}\,\textsl{\texttt C}_g^-}
{64 \pi^2\,|\varepsilon_{\rm exp}| f\!_\pi^{}\, {\rm Re}A_0^{\rm exp}}
\,=\, (-1.4{\pm}0.4) \!\times\! 10^5\, {\rm GeV~Im}\,\textsl{\texttt C}_g^- , ~
\end{align}
with \,$\omega = {\rm Re}\,A_2^{\rm exp}/{\rm Re}\,A_0^{\rm exp}$\, involving the empirical values~\cite{Cirigliano:2011ny}
\,Re\,$A_0^{\rm exp} =  (2.704\pm0.001) \times10^{-7} \rm\,GeV$  and \,Re\,$A_2^{\rm exp} = (1.210 \pm 0.002) \times10^{-8} \rm\,GeV$.\,
Based on the 2$\sigma$ range of the difference between the data~\cite{ParticleDataGroup:2024cfk} \,$(\varepsilon'/\varepsilon)_{\rm exp}=(16.6\pm2.3)\times10^{-4}$\, and the SM prediction~\cite{Cirigliano:2019cpi} \,$(\varepsilon'/\varepsilon)_{\tt SM}^{}=(14\pm5)\times10^{-4}$,\, we may impose \,$-0.8 < 10^3\, \varepsilon_{\tt CMO}'/\varepsilon < 1.4$,\, which translates into
\begin{equation} \label{ImCg-bound}
-9.7\times10^{-9} {\rm\,GeV^{-1}} \,<\, {\rm Im}\,\textsl{\texttt C}_g^- \,<\, 6.0\times10^{-9} \rm\,GeV^{-1} \,. ~~~
\end{equation}

In evaluating the hyperon $CP$-asymmetries amplified by new physics via the CMOs, we take into account that the $\xi$s in eq.\,(\ref{xi-cmo}) have relative uncertainties of $\cal O$(1) and Im$\,\textsl{\texttt C}_g^\pm$ are subject to the restrictions in eqs.\,\,(\ref{ImCg+bound}) and~(\ref{ImCg-bound}).
Accordingly, incorporating the central values of the components in eq.\,(\ref{xAxB}) and of the strong phases in eq.\,(\ref{deltas}), we accumulate $10^5$ sets of the asymmetries from weak phases which are randomly generated from a uniform distribution and fulfill the preceding requisites.
We then arrive at the following averages and their 1$\sigma$ errors:
\begin{align} \label{cpa-cmo}
\hat A_{\mathit\Sigma^+\to n\pi^+}^{\tt CMO} & \,=\, (4.0\pm7.1) \times10^{-3} \,, &
\hat B_{\mathit\Sigma^+\to n\pi^+}^{\tt CMO} & \,=\, (0.0\pm3.6) \times10^{-2} \,,
\nonumber \\
\hat A_{\mathit\Sigma^+\to p\pi^0}^{\tt CMO} & \,=\, (0.6\pm1.0) \times10^{-4} \,, &
\hat B_{\mathit\Sigma^+\to p\pi^0}^{\tt CMO} & \,=\, (-0.7\pm1.4) \times10^{-3} \,, ~~~~~
\nonumber \\
\hat A_{\mathit\Sigma^-\to n\pi^-}^{\tt CMO} & \,=\, (0.1\pm2.2) \times10^{-3} \,, &
\hat B_{\mathit\Sigma^-\to n\pi^-}^{\tt CMO} & \,=\, (-0.1\pm1.7) \times10^{-2} \,,
\nonumber \\
\Delta\phi_{\mathit\Sigma^+\to n\pi^+}^{\tt CMO} & \,=\, (0.0\pm1.8) \times 10^{-3} \,,  &
\hat\Delta_{\mathit\Sigma^+\to n\pi^+}^{\tt CMO} & \,=\, (0.1\pm2.3) \times 10^{-4} \,,
\nonumber \\
\Delta\phi_{\mathit\Sigma^+\to p\pi^0}^{\tt CMO} & \,=\, (3.8\pm7.7) \times 10^{-3} \,,  &
\hat\Delta_{\mathit\Sigma^+\to p\pi^0}^{\tt CMO} & \,=\, (-0.1\pm2.3) \times 10^{-4} \,,
\nonumber \\
\Delta\phi_{\mathit\Sigma^-\to n\pi^-}^{\tt CMO} & \,=\, (0.1\pm1.1) \times 10^{-3} \,.
\end{align}
Like the SM instances, $\hat A_{\mathit\Sigma^+\to n\pi^+}^{\tt CMO}$ and $\hat A_{\mathit\Sigma^+\to p\pi^0}^{\tt CMO}$ agree with their empirical counterparts in eq.\,(\ref{ASigma-data}) at the 2$\sigma$ level, as can be viewed in figure~\ref{A-exp-th}.

The chromomagnetic-penguin interactions influence other nonleptonic hyperon decays, such as \,$\mathit\Lambda\to N\pi$\, and \,$\mathit\Xi\to\mathit\Lambda\pi$.\,
For completeness, appendix~\ref{LXcpv-cmo} contains the predictions for $CP$-odd signals similarly magnified by the CMOs that can be probed with these $\mathit\Lambda$ and $\mathit\Xi$ channels.

\subsection{Enhanced contributions in dark-matter model\label{dmmodel}}

In this subsection we consider the model adopted in ref.\,\cite{He:2025jfc} and named THDM+D, which is the two-Higgs-doublet model of type III plus a real scalar particle $D$ acting as a dark-matter candidate.
This scenario not only provides the quark transition \,$b\to sDD$\, mediated at tree level by the heavy $CP$-even Higgs boson $H$ and inspired by recent measurements~\cite{Belle-II:2023esi} of the $b$-meson decay \,$B^+\to K^+\nu\bar\nu$,\, but also can satisfy the relic-density requirement and limits from direct and indirect searches for dark matter.
Another important feature of the THDM+D is that it can supply new four-quark interactions, caused by tree-level exchanges of the heavy Higgses, which bring about significant $CP$-violation in \,$\mathit\Lambda\to N\pi$\, and \,$\mathit\Xi\to\mathit\Lambda\pi$,\, as demonstrated in ref.\,\cite{He:2025jfc}.
It is of interest to see if this could likewise occur in \,$\mathit\Sigma\to N\pi$.\,

The relevant quark interactions are described by the effective Lagrangian~\cite{He:2025jfc}
\begin{equation} \label{L4q}
{\cal L}_{4q}^{\rm new} \,\supset\, {\cal C}_u^{} {\cal Q}_u^{} + {\cal C}_+^{} {\cal Q}_+^{} + {\cal C}_-^{} {\cal Q}_-^{} \,+\, \rm H.c. \,, ~~~
\end{equation}
where
\begin{align} \label{C-+u}
{\cal C}_u^{} & \,=\, V_{us}^{} \mathbb Y_{ss}^{}\, \frac{ V_{ud}^* \mathbb Y_{dd}^* + V_{us}^* \mathbb Y_{sd}^* }{m_H^2} \,, & {\cal C}_\pm^{} & \,=\, \mathbb Y_{sd}^*\, \frac{\mathbb Y_{dd}^{}\pm\mathbb Y_{ss}^{}}{2 m_H^2} \,\equiv\, \frac{{\cal C}_d\pm{\cal C}_s}{2} \,, ~~~ ~~
\nonumber \\
{\cal Q}_u^{} & \,=\, \overline{u_L^{}}s_R^{}\, \overline{d_R^{}}u_L^{} \,, & {\cal Q}_\pm^{} & \,=\, \overline{d_R^{}}s_L^{}\, \big( \overline{d_L^{}}d_R^{} \pm \overline{s_L^{}}s_R^{} \big) \,.
\end{align}
Here $\mathbb Y_{dd,ds,ss}$ are generally complex Yukawa couplings and hence constitute additional sources of $CP$ violation, $m_H$ is the $H$ mass, and the heavy Higgses have been assumed to possess the same mass of~1~TeV.

To address the impact of ${\cal L}_{4q}^{\rm new}$ on hyperon and kaon processes, we need the hadronic realization of eq.\,(\ref{L4q}), which must share the symmetry properties exhibited therein.
At leading order it has been derived in ref.\,\cite{He:2025jfc} to be
\begin{align} \label{Lchi} &
{\cal L}_\chi^{\rm new} \,\supset~ \tilde{\textsc c}_u^{}\, {\cal O}_u^{} + \tilde{\textsc c}_+^{}\, {\cal O}_+^{} + \tilde{\textsc c}_-^{}\, {\cal O}_-^{} \,+\, \rm H.c. \,, &
\end{align}
where
\begin{align} \label{cuc+-}
\tilde{\textsc c}_u^{} \,= & ~ \hat\eta_u^{}\, {\cal C}_u^{} \,, ~~~ ~~ \hat\eta_u^{} \,=\, 4.7 \,, ~~~ ~~~~ ~~
\tilde{\textsc c}_\pm^{} \,=\, \tfrac{1}{2}\, \hat\eta_d^{}\,  \big({\cal C}_d^{} \pm {\cal C}_s^{} \big) \,, ~~~ ~~ \hat\eta_d^{}  \,=\, 4.8 \,, ~~~ ~~~~
\end{align}
the factors $\hat\eta_{u,d}^{}$ appearing due to QCD running from 1 TeV to 1 GeV, and
\begin{align} \label{Qu}
{\cal O}_u^{} & \,=\, \hat{\textsc d}_u^{}
\Big[ \big( \xi^\dagger \big\{ B, \,\overline{\!B} \big\} \xi^\dagger \big)_{\!31} \hat\Sigma_{12}^{}
\!+\! \hat\Sigma_{31}^\dagger \big( \xi \big\{ B, \,\overline{\!B} \big\} \xi \big)_{\!12} \Big]
\!+\! \hat{\textsc f}\!_u^{} \Big[ \big( \xi^\dagger \big[ B, \,\overline{\!B} \big] \xi^\dagger \big)_{\!31}
\hat\Sigma_{12}^{} \!+\! \hat\Sigma_{31}^\dagger \big( \xi \big[ B, \,\overline{\!B} \big] \xi \big)_{\!12} \Big] ~
\nonumber \\ & ~~~~ +\, \hat{\textsc g}_u^{}\, \big( \xi^\dagger \,\overline{\!B} \xi^\dagger \big)_{31} (\xi B\xi)_{12}^{} + \hat{\textsc g}{}_u'\, \big(\xi \,\overline{\!B} \xi\big)_{12} \big( \xi^\dagger B \xi^\dagger \big)_{31}
+ \hat{\textsc h}_u^{}\, \hat\Sigma_{31}^\dagger \hat\Sigma_{12}^{} \,,
\end{align}
\begin{align} \label{Qpm}
{\cal O}_\pm^{} \,= & ~ \Big[  \hat{\textsc d}_\pm^{}\, \big( \xi \big\{ B, \,\overline{\!B} \big\} \xi \big)_{32}
+ \hat{\textsc f}_\pm^{}\, \big( \xi \big[ B, \,\overline{\!B} \big] \xi \big)_{32} \Big] \big( \hat\Sigma_{22}^\dagger \pm \hat\Sigma_{33}^\dagger \big)
\nonumber \\ & +\, \hat\Sigma_{32}^{}\, \Big[ \hat{\textsc d}{}_\pm'\,
\big( \xi^\dagger \big\{ B, \,\overline{\!B} \big\} \xi^\dagger \big)_{22} + \hat{\textsc f}{}_\pm'\,
\big( \xi^\dagger \big[ B, \,\overline{\!B} \big] \xi^\dagger \big)_{22} \pm (2{\,\to\,}3) \Big]
\nonumber \\ & +\,
\hat{\textsc g}_\pm^{} \Big\{ \big(\xi \,\overline{\!B} \xi\big)_{32} \Big[ \big( \xi^\dagger B \xi^\dagger \big)_{22} \pm  (2{\,\to\,}3) \Big] + \Big[ \big( \xi^\dagger \,\overline{\!B} \xi^\dagger \big)_{22}
\pm (2{\,\to\,}3) \Big] \big(\xi B\xi\big)_{32} \Big\}
\nonumber \\ & +\, \hat{\textsc h}_\pm^{}\, \hat\Sigma_{32}^{} \big( \hat\Sigma_{22}^\dagger\pm\Sigma_{33}^\dagger \big) \,, &
\end{align}
with \cite{He:2025jfc}
\begin{align} \label{fact}
\hat{\textsc d}_{u,-}^{} & \,=\, \hat{\textsc d}{}_\pm' \,=\, 0.00184\rm~GeV^3 \,,  &
\hat{\textsc d}_+^{} & \,=\, 0.00199\rm~GeV^3 \,,
\nonumber \\
\hat{\textsc f}_{u,-}^{} & \,=\, \hat{\textsc f}{}_\pm' \,=\, -0.00660\rm~GeV^3 \,, &
\hat{\textsc f}_+^{} & \,=\, -0.00645\rm~GeV^3 \,,
\nonumber \\
\hat{\textsc g}_u^{} & \,=\, -3.45\times 10^{-4}\rm~GeV^3 \,, & \hat{\textsc g}_\pm^{} & \,=\, -2.97\times 10^{-4}\rm~GeV^3 \,, & \hat{\textsc g}{}_u' & \,=\, 0 \,,  ~~~ ~~
\nonumber \\
\hat{\textsc h}_u^{} & \,=\, 6.49 \times 10^{-5} \rm~GeV^6 \,, & \hat{\textsc h}_\pm^{} & \,=\, 6.62 \times 10^{-5} \rm~GeV^6 \,.
\end{align}

The contributions of ${\cal L}_\chi^{\rm new}$ to $\mathbb A$ and $\mathbb B$ at leading order are calculated from, respectively, the contact and tadpole diagrams depicted in figure\,\,\ref{npdiagrams}(a) and the baryon- and kaon-pole diagrams in figure\,\,\ref{npdiagrams}(b).
Thus, we find
\begin{align} \label{newA}
\mathbb A_{\mathit\Sigma^+\to n\pi^+}^{\rm new} & \,=\, \frac{ -\widetilde{\textsc c}_-^{}
\hat{\textsc g}_-^{} - \widetilde{\textsc c}_+^{} \hat{\textsc g}_+^{} - \widetilde{\textsc c}_u^{} \hat{\textsc g}_u^{} }{\sqrt2\,f_\pi^{}} \,,
\nonumber \\
\mathbb A_{\mathit\Sigma^+\to p\pi^0}^{\rm new} & \,=\, \frac{ 2 \widetilde{\textsc c}_-^{}\, \big( \hat{\textsc d}_-^{} - \hat{\textsc f}_-^{} \big) - \widetilde{\textsc c}_u^{} \hat{\textsc g}_u^{} }{2f_\pi^{}} + \frac{\widetilde{\textsc c}_+^{} \hat{\textsc h}_+^{}}{f_\pi^3} \bigg(
\frac{m_N^{}-m_\mathit\Sigma^{}}{m_\pi^2-m_K^2} \bigg) \,,
\nonumber \\
\mathbb A_{\mathit\Sigma^-\to n\pi^-}^{\rm new} & \,=\, \frac{ \widetilde{\textsc c}_+^{} \big( 2 \hat{\textsc d}_+^{} - 2 \hat{\textsc f}_+^{} - \hat{\textsc g}_+^{} \big)
- \widetilde{\textsc c}_-^{} \hat{\textsc g}_-^{} + 2 \widetilde{\textsc c}_u^{} \big(
\hat{\textsc d}_u^{} - \hat{\textsc f}_u^{} \big) }{\sqrt2\,f_\pi^{}} + \sqrt2\, \frac{\widetilde{\textsc c}_+^{} \hat{\textsc h}_+^{}\, (m_\mathit\Sigma-m_N)}{f_\pi^3~ \big(m_\pi^2-m_K^2\big)} \,,
\end{align}
\begin{align} \label{newB}
\mathbb B_{\mathit\Sigma^+\to n\pi^+}^{\rm new} & \,=\, \frac{m_\mathit\Sigma^{}+m_N^{}}{\sqrt2\,f_\pi^{}} \!\! \begin{array}[t]{l} \displaystyle \Bigg\{ \Bigg[ \frac{ 3\, \widetilde{\textsc c}_-^{} \hat{\textsc g}_-^{} - \widetilde{\textsc c}_+^{}\, \big( 2\, \hat{\textsc d}_+^{} + 6\, \hat{\textsc f}_+^{} + \hat{\textsc g}_+^{} \big) }
{m_\mathit\Lambda^{}-m_N^{}} \Bigg] \frac{\cal D}{3}
\\ \displaystyle ~+\, \frac{ 2 \widetilde{\textsc c}_+^{} \big( \hat{\textsc d}_+^{} - \hat{\textsc f}_+^{} \big) + \widetilde{\textsc c}_u^{} \hat{\textsc g}_u^{} }{m_N^{}-m_\mathit\Sigma^{}}\, {\cal D} + \frac{ \widetilde{\textsc c}_-^{} \hat{\textsc g}_-^{}
+ \widetilde{\textsc c}_+^{} \hat{\textsc g}_+^{} - \widetilde{\textsc c}_u^{}
\hat{\textsc g}_u^{} }{m_\mathit\Sigma^{}-m_N^{}}\, {\cal F} \Bigg\} \,, \end{array} &
\nonumber \\
\mathbb B_{\mathit\Sigma^+\to p\pi^0}^{\rm new} & \,=\, \frac{m_\mathit\Sigma^{}+m_N^{}}{f_\pi^{}} \Bigg[ \frac{ 2\, \widetilde{\textsc c}_+^{}\, \big(\hat{\textsc d}_+^{}-\hat{\textsc f}_+^{}\big) + \widetilde{\textsc c}_u^{} \hat{\textsc g}_u^{} }{2\big(m_N^{}-m_\mathit\Sigma^{}\big)} + \frac{\widetilde{\textsc c}_-^{}\hat{\textsc h}_-^{}}{f_\pi^2~\big(m_\pi^2-m_K^2\big)} \Bigg] ({\cal D-F}) \,,
\nonumber \\
\mathbb B_{\mathit\Sigma^-\to n\pi^-}^{\rm new} & \,=\, \frac{m_\mathit\Sigma^{}+m_N^{}}{\sqrt2\,f_\pi^{}} \!\!
\begin{array}[t]{l} \displaystyle \Bigg\{ \Bigg[ \frac{ 3\, \widetilde{\textsc c}_-^{}
\hat{\textsc g}_-^{} - \widetilde{\textsc c}_+^{}\, \big( 2\, \hat{\textsc d}_+^{}
+ 6\, \hat{\textsc f}_+^{} + \hat{\textsc g}_+^{} \big) }{m_\mathit\Lambda^{}-m_N^{}} \Bigg] \frac{\cal D}{3}
\\ \displaystyle ~+\, \frac{ \widetilde{\textsc c}_-^{}\hat{\textsc g}_-^{} + \widetilde{\textsc c}_+^{}\, \big(2\,\hat{\textsc d}_+^{} - 2\,\hat{\textsc f}_+^{} + \hat{\textsc g}_+^{}\big) }{m_N^{}-m_\mathit\Sigma^{}}\, {\cal F}
\\ \displaystyle ~+\, 2 \Bigg( \frac{ \widetilde{\textsc c}_+ \hat{\textsc h}_+ - \widetilde{\textsc c}_u \hat{\textsc h}_u }{m_\pi^2-m_K^2} \Bigg) \frac{{\cal D}-\cal F}{f_\pi^2}\Bigg\} \,. \end{array}
\end{align}
From these, it is straightforward to derive their isospin components $\dot{\tt A}_1^{\rm new}$, $\dot{\tt A}_{1,3}^{\rm new}$, $\dot{\tt A}_{3,3}^{\rm new}$, $\dot{\tt B}_1^{\rm new}$, $\dot{\tt B}_{1,3}^{\rm new}$, and $\dot{\tt B}_{3,3}^{\rm new}$, in analogy to the experimental ones in eq.\,(\ref{xAxB}), in order
to compute the weak phases induced by ${\cal L}_\chi^{\rm new}$.

To explore how this affects the $CP$-violating observables in \,$\mathit\Sigma\to N\pi$,\, we employ the sample values of Yukawa couplings ($\mathbb Y_{dd,ds,ss}$) permitted by dark-matter and kaon constraints which were obtained in ref.\,\cite{He:2025jfc}.
Since the real parts of $\mathbb A$ and $\mathbb B$ may be enlarged by these new contributions, and since as explained in section~\ref{smpredict} the theoretical treatment of $\mathit\Sigma\to N\pi$ amplitudes suffers from significant uncertainties, here we impose an extra condition that $|\dot{\tt A}_x^{\rm new}|$ and $|\dot{\tt B}_x^{\rm new}|$ be less than 25\% of their empirical counterparts in eq.\,(\ref{xAxB}).

\begin{figure}[b] \medskip \centering
\includegraphics[width=0.31\linewidth]{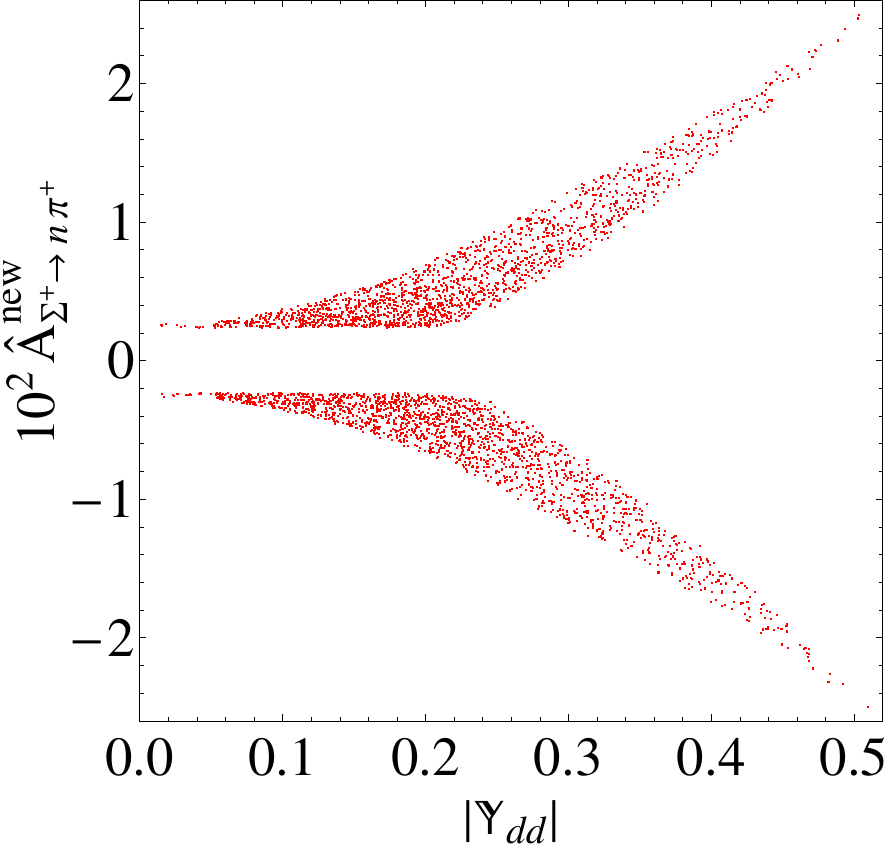} ~~ \includegraphics[width=0.31\linewidth]{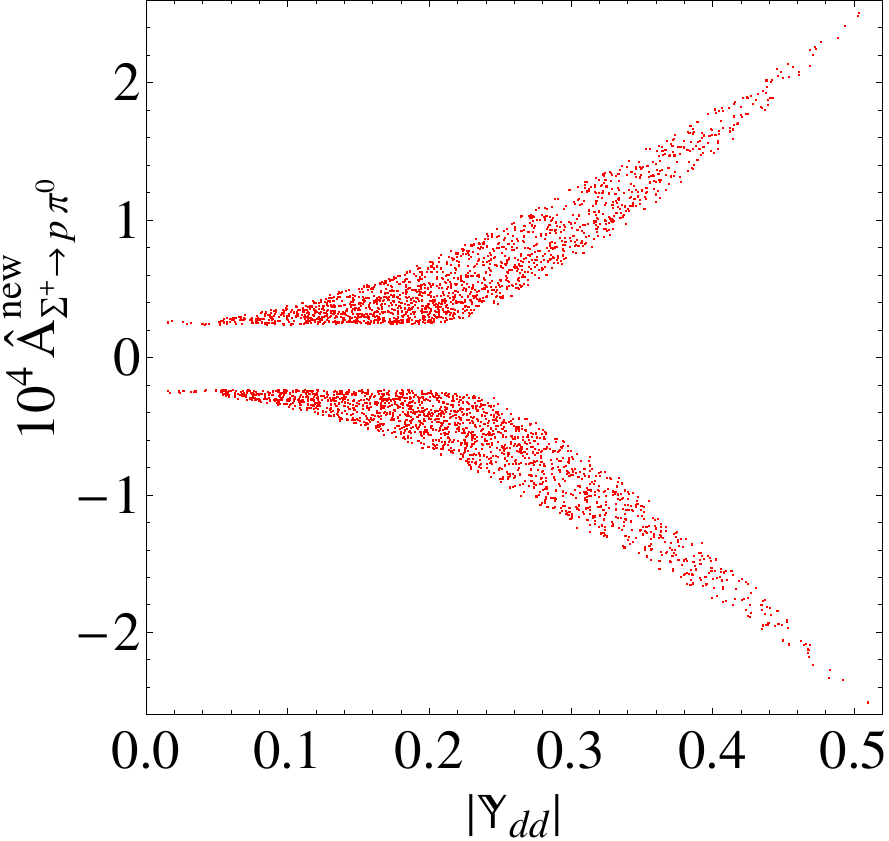} ~~ \includegraphics[width=0.31\linewidth]{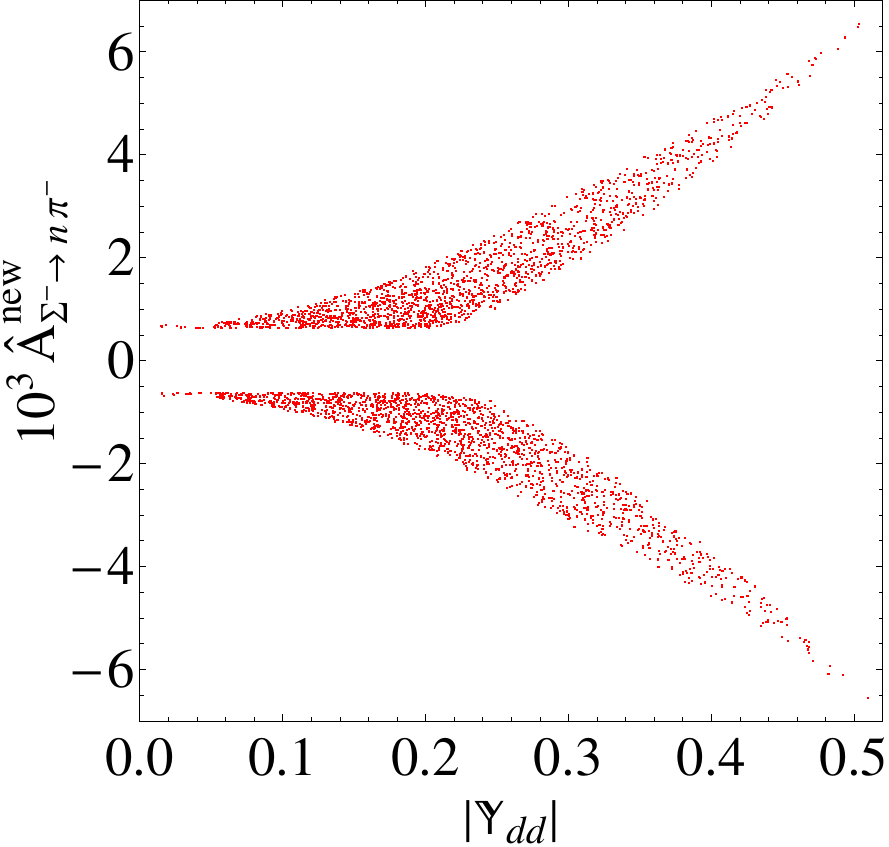}\vspace{1ex}\\
\includegraphics[width=0.31\linewidth]{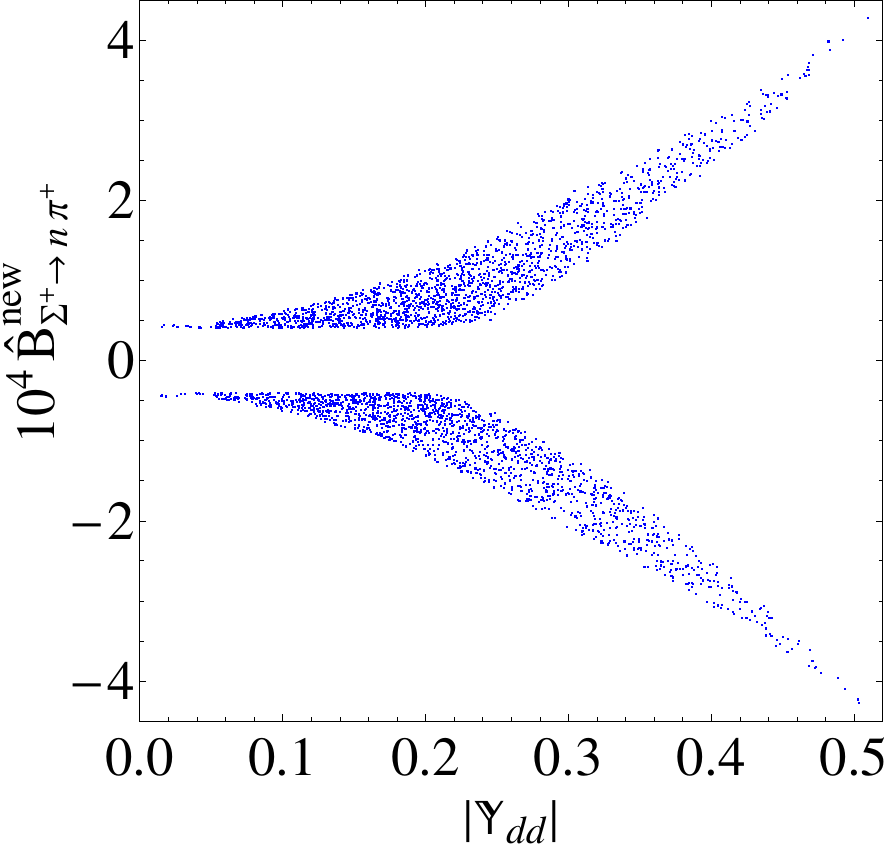} ~~ \includegraphics[width=0.31\linewidth]{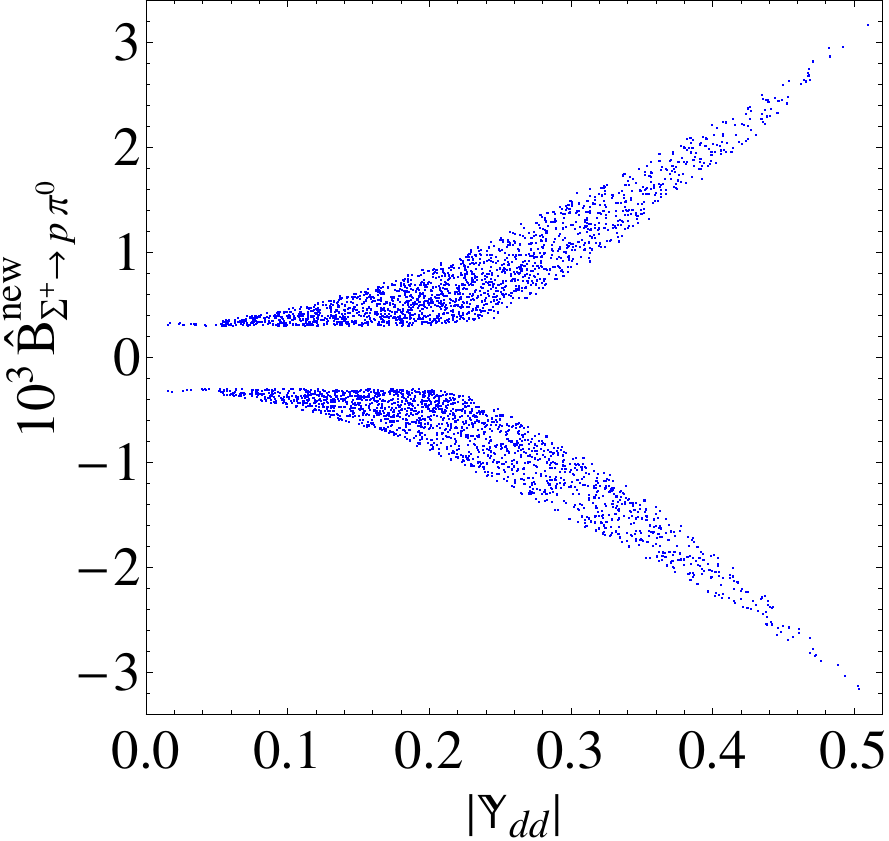} ~~ \includegraphics[width=0.31\linewidth]{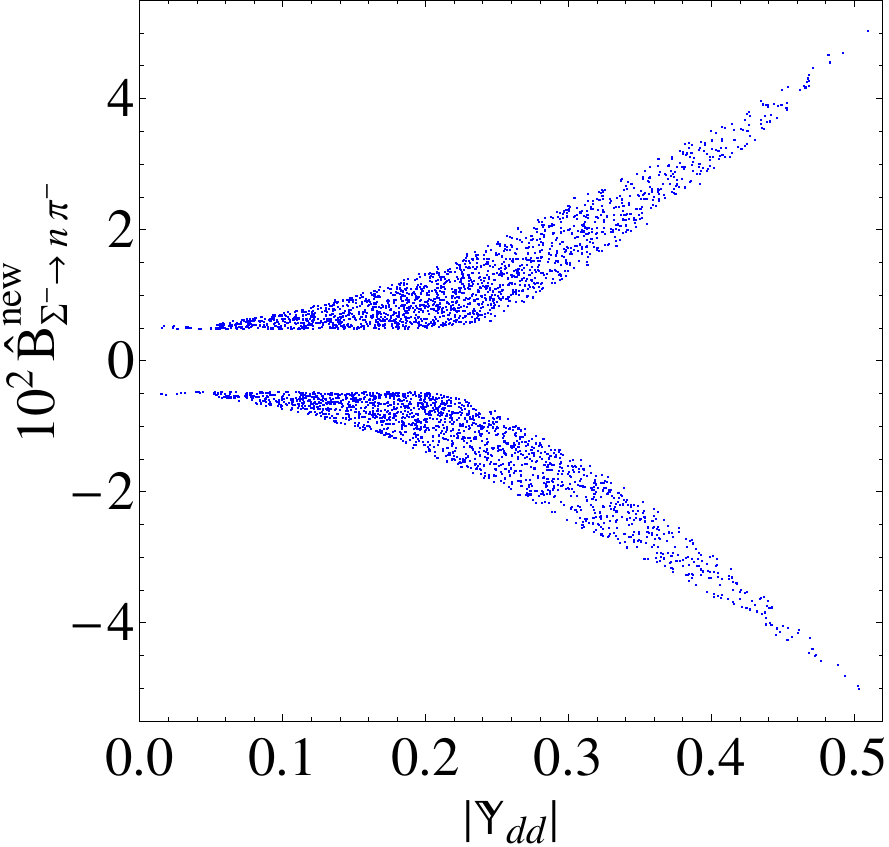}\vspace{-2pt}
\caption{Top: the distributions of $CP$ asymmetries $\hat A_{\mathit\Sigma^+\to n\pi^+}^{\rm new}$, $\hat A_{\mathit\Sigma^+\to p\pi^0}^{\rm new}$, and $\hat A_{\mathit\Sigma^-\to n\pi^-}^{\rm new}$ in connection with the sample absolute-values of Yukawa coupling $\mathbb Y_{dd}$ fulfilling the requirements specified in the text.
Bottom: the corresponding $\hat B_{\mathit\Sigma^+\to n\pi^+}^{\rm new}$, $\hat B_{\mathit\Sigma^+\to p\pi^0}^{\rm new}$, and $\hat B_{\mathit\Sigma^-\to n\pi^-}^{\rm new}$.} \label{2hdmd-plots}
\end{figure}

The resulting allowed Yukawa couplings and the central values of the various input parameters translate into the $CP$-asymmetry ranges
\begin{align} \label{cpa-thdmd}
-2.5\times10^{-2} & \le \hat A_{\mathit\Sigma^+\to n\pi^+}^{\rm new} \le 2.5\times10^{-2} \,, &
-4.3\times10^{-4} & \le \hat B_{\mathit\Sigma^+\to n\pi^+}^{\rm new} \le 4.3\times10^{-4} \,, &
\nonumber \\
-2.5\times10^{-4} & \le \hat A_{\mathit\Sigma^+\to p\pi^0}^{\rm new} \le 2.5\times10^{-4} \,, &
-3.2\times10^{-3} & \le \hat B_{\mathit\Sigma^+\to p\pi^0}^{\rm new} \le 3.2\times10^{-3} \,, &
\nonumber \\
-6.6\times10^{-3} & \le \hat A_{\mathit\Sigma^-\to n\pi^-}^{\rm new} \le 6.5\times10^{-3} \,, &
-5.0\times10^{-2} & \le \hat B_{\mathit\Sigma^-\to n\pi^-}^{\rm new} \le 5.0\times10^{-2} \,,
\nonumber \\
-6.1\times10^{-6} & \le \Delta\phi_{\mathit\Sigma^+\to n\pi^+}^{\rm new} \le
6.1\times10^{-6} \,, &
-2.2\times10^{-5} & \le \hat \Delta_{\mathit\Sigma^+\to n\pi^+}^{\rm new} \le
2.2\times10^{-5} \,,
\nonumber \\
-1.7\times10^{-2} & \le \Delta\phi_{\mathit\Sigma^+\to p\pi^0}^{\rm new} \le
1.7\times10^{-2} \,, &
-2.4\times10^{-5} & \le \hat\Delta_{\mathit\Sigma^+\to p\pi^0}^{\rm new} \le
2.4\times10^{-5} \,,
\nonumber \\
-3.4\times10^{-3} & \le \Delta\phi_{\mathit\Sigma^-\to n\pi^-}^{\rm new} \le
3.4\times10^{-3} \,.
\end{align}
For illustration, the distributions of $\hat A_{\mathit\Sigma\to N\pi}^{\rm new}$ and $\hat B_{\mathit\Sigma\to N\pi}^{\rm new}$ for the three modes versus $|\mathbb Y_{dd}|$ are depicted in figure~\ref{2hdmd-plots}, where asymmetries with magnitudes less than {\small\,$\sim$\,}10\% of their respective maxima are not displayed.
Their corresponding relations to $|\mathbb Y_{sd,ss}|$ are visually alike and hence not shown.
It is evident that the relative magnitude of $\hat A_{\mathit\Sigma^+\to n\pi^+}^{\rm new}$ and $\hat B_{\mathit\Sigma^+\to n\pi^+}^{\rm new}$ is different from those in the other two modes where \,$|\hat A|\sim0.1\,|\hat B|$.\, 
Such situations appear to arise as well in the SM and CMO instances in eqs.\,\,(\ref{cpa-sm}) and (\ref{cpa-cmo}), respectively, which were also found in the past~\cite{Donoghue:1986hh}.\footnote{By contrast, for all of the \,$\mathit\Lambda\to N\pi$\, and \,$\mathit\Xi\to\mathit\Lambda\pi$\, modes \,$|\hat A|\sim0.1\,|\hat B|$\, or less within the SM and beyond, as reported before~\cite{Donoghue:1985ww,Donoghue:1986hh}. This holds even with exact formulas for the asymmetries, as indicated by the CMO examples in eqs.\,\,(\ref{cpa-L-cmo}) and~(\ref{cpa-X-cmo}).}

The $\hat A_{\mathit\Sigma\to N\pi}^{\rm new}$ results in eq.\,(\ref{cpa-thdmd}) are bigger than the SM expectations in eq.\,(\ref{cpa-sm}) by about 10 times, but unlike the CMO case the THDM+D contributions to a few of the other asymmetries are not.
Such dissimilarities of the different scenarios are potentially testable in future experiments.

For comparison with the data in eq.\,(\ref{ASigma-data}), we have drawn in figure\,\,\ref{A-exp-th} the blue-colored areas labeled \textsf{4q} depicting the spans of $\hat A_{\mathit\Sigma^+\to n\pi^+}^{\rm new}$ and $\hat A_{\mathit\Sigma^+\to p\pi^0}^{\rm new}$, after including the effects of the $\cal O$(1) relative uncertainties of the weak phases.
As can be viewed in the figure, $\hat A_{\mathit\Sigma^+\to n\pi^+}^{\rm new}$ overlaps with the 1$\sigma$ interval of its empirical counterpart, whereas $\hat A_{\mathit\Sigma^+\to p\pi^0}^{\rm new}$ agree with its data at 2$\sigma$.

Needless to say, what figure\,\,\ref{A-exp-th} reveals invites more precise measurements of these quantities, which will scrutinize our predictions more stringently.
This adds to the importance of the proposed Super Tau Charm Facility~\cite{Achasov:2023gey}, where measurements of $\hat A$ and $\hat B$ for \,$\mathit\Sigma\to N\pi$,\, as well as for \,$\mathit\Lambda\to N\pi$\, and \,$\mathit\Xi\to\mathit\Lambda\pi$,\, are anticipated to have statistical precisions reaching order $10^{-4}$ or better~\cite{Salone:2022lpt,Guo:2025bfn}.

\section{Conclusions\label{conclusions}}

Motivated in part by recent and upcoming BESIII measurements of the $\mathit\Sigma^\pm$ hyperons, we have carried out a theoretical study of several quantities that can test for $CP$ violation in \,$\mathit\Sigma^\pm\to n\pi^\pm$  and \,$\mathit\Sigma^+\to p\pi^0$\, decays, which has not been done in the recent literature.
After addressing these $CP$-odd observables within the standard model, incorporating the relevant up-to-date information, we perform the corresponding analyses in a couple of new-physics scenarios that could produce substantial $CP$-violation in the \,$\mathit\Lambda\to N\pi$\, and \,$\mathit\Xi\to\mathit\Lambda\pi$\, channels.
In the first scenario, we examine in a model-independent manner the impact of amplified chromomagnetic-penguin interactions, taking into account constraints from the kaon sector.
Our numerical work demonstrates that the $CP$ asymmetries in the $\Sigma^\pm$ modes can be larger than their respective standard-model expectations by up to an order of magnitude.
We arrive at comparable conclusions in the second scenario, which is a particular model containing two Higgs doublets and in which new scalar four-quark operators arising from tree-level exchange of heavy Higgs bosons can yield magnified effects.
Our results concerning sizable $CP$-violation in \,$\mathit\Sigma\to N\pi$\, can potentially be probed by near-future experiments.
As another incentive for such efforts, it is interesting to mention that the current data on $\hat A_{\mathit\Sigma^+\to n\pi^+}$ and $\hat A_{\mathit\Sigma^+\to p\pi^0}$ are approaching their predictions in the SM and in the new-physics cases considered, a situation which may be clarified with additional measurements.

\section*{Acknowledgements}

We thank Hong-Fei Shen and Xiaorong Zhou for experimental information.
J.T. and G.V. thank the Tsung-Dao Lee Institute, Shanghai Jiao Tong University, for kind hospitality and support during the early stages of this research.
X.-G.H. was supported by the Fundamental Research Funds for the Central Universities, by the National Natural Science Foundation of the People's Republic of China (Nos. 12090064, 11735010, 12205063, 11985149, 12375088, and W2441004), and by MOST 109-2112-M-002-017-MY3. X.-D.M. was supported by Grant No. NSFC-12305110.

\appendix

\section{Decay parameters of \boldmath$\mathit\Sigma^-\to n\pi^-$ and $\,\overline{\!\mathit\Sigma}{}^+\to\overline n\pi^+$\label{app:Sigma-}}

Since the final states of \,$\mathit\Sigma^-\to n\pi^-$\, and its antiparticle counterpart, \,$\overline{\!\mathit\Sigma}{}^+\to\overline n\pi^+$,\, have a total isospin with just one value, \,$I_{\rm f}=3/2$,\, there is only one strong phase in each of their $\tt S$- and $\tt P$-wave components, $\delta_3^{\rm S}$ and $\delta_3^{\rm P}$, respectively.
The $\mathbb A$ and $\mathbb B$ parts of their amplitudes are therefore expressible as~\cite{Donoghue:1986hh,Overseth:1969bxc}
\begin{align} \label{AB-S}
\mathbb A_{\mathit\Sigma^-\to n\pi^-}^{} & =\, \bigg( {\tt A}_{1,3}^{}\, e^{i\xi_{13}^{\rm S}} + \sqrt{\tfrac{2}{5}}~ {\tt A}_{3,3}^{}\, e^{i\xi_{33}^{\rm S}} + \tfrac{1}{\sqrt{15}}\, {\tt A}_{5,3}^{}\, e^{i\xi_{53}^{\rm S}} \bigg) e^{i\delta_3^{\rm S}} \,,
\nonumber \\
\mathbb B_{\mathit\Sigma^-\to n\pi^-}^{} & =\, \bigg( {\tt B}_{1,3}^{}\, e^{i\xi_{13}^{\rm P}} + \sqrt{\tfrac{2}{5}}~ {\tt B}_{3,3}^{}\, e^{i\xi_{33}^{\rm P}} + \tfrac{1}{\sqrt{15}}\, {\tt B}_{5,3}^{}\, e^{i\xi_{53}^{\rm P}} \bigg) e^{i\delta_3^{\rm P}} \,,
\nonumber \\
\mathbb A_{\,\overline{\!\mathit\Sigma}{}^+\to\overline n\pi^+}^{} & \,=\, -\bigg( {\tt A}_{1,3}^{}\, e^{-i\xi_{13}^{\rm S}} + \sqrt{\tfrac{2}{5}}~ {\tt A}_{3,3}^{}\, e^{-i\xi_{33}^{\rm S}} + \tfrac{1}{\sqrt{15}}\, {\tt A}_{5,3}^{}\, e^{-i\xi_{53}^{\rm S}} \bigg) e^{i\delta_3^{\rm S}} \,,
\nonumber \\
\mathbb B_{\,\overline{\!\mathit\Sigma}{}^+\to\overline n\pi^+}^{} & =\, +\bigg( {\tt B}_{1,3}^{}\, e^{-i\xi_{13}^{\rm P}} + \sqrt{\tfrac{2}{5}}~ {\tt B}_{3,3}^{}\, e^{-i\xi_{33}^{\rm P}} + \tfrac{1}{\sqrt{15}}\, {\tt B}_{5,3}^{}\, e^{-i\xi_{53}^{\rm P}} \bigg) e^{i\delta_3^{\rm P}} \,, ~~~~~
\end{align}
where ${\tt A}_{2\Delta I,3}$ and ${\tt B}_{2\Delta I,3}$ on the right-hand sides are real and associated with the isospin changes \,$\Delta I=1/2,3/2,5/2$\, caused by the transitions.

From eq.\,(\ref{AB-S}), it follows that
\begin{align} \label{|ASigma-|}
|\mathbb A_{\mathit\Sigma^-\to n\pi^-}| & \,=\, |\mathbb A_{\,\overline{\!\mathit\Sigma}{}^+\to\overline n\pi^+}| \,, & |\mathbb B_{\mathit\Sigma^-\to n\pi^-}| & \,=\, |\mathbb B_{\,\overline{\!\mathit\Sigma}{}^+\to\overline n\pi^+}| \,, ~~~ ~~
\end{align}
and consequently
\begin{align} \label{gammabar}
\Gamma_{\mathit\Sigma^-\to n\pi^-}^{} & \,=\, \Gamma_{\,\overline{\!\mathit\Sigma}{}^+\to\overline n\pi^+} \,, &
\gamma_{\mathit\Sigma^-\to n\pi^-}^{} & \,=\, \gamma_{\,\overline{\!\mathit\Sigma}{}^+\to\overline n\pi^+}^{} \,, ~~~ ~~~~
\end{align}
the second relation implying
\begin{align} \label{a^2}
\alpha_{\mathit\Sigma^-\to n\pi^-}^2 + \beta_{\mathit\Sigma^-\to n\pi^-}^2 & \,=\, \alpha_{\,\overline{\!\mathit\Sigma}{}^+\to\overline n\pi^+}^2 + \beta_{\,\overline{\!\mathit\Sigma}{}^+\to\overline n\pi^+}^2 \,. ~~~~~~~
\end{align}
Moreover, from eq.\,(\ref{AB-S}), it is straightforward to derive
\begin{equation} \label{ratios}
\frac{\beta_{\mathit\Sigma^-\to n\pi^-} - \beta_{\,\overline{\!\mathit\Sigma}{}^+\to\overline n\pi^+}}{\alpha_{\mathit\Sigma^-\to n\pi^-}^{} - \alpha_{\,\overline{\!\mathit\Sigma}{}^+\to\overline n\pi^+}^{}} \,=\, \frac{-\alpha_{\mathit\Sigma^-\to n\pi^-}^{} - \alpha_{\,\overline{\!\mathit\Sigma}{}^+\to\overline n\pi^+}^{}}{\beta_{\mathit\Sigma^-\to n\pi^-} + \beta_{\,\overline{\!\mathit\Sigma}{}^+\to\overline n\pi^+}} \,=\, \tan\big(\delta_3^{\rm P}-\delta_3^{\rm S}\big) \,, ~~~ ~~
\end{equation}
where the amplitudes and weak phases have completely dropped out from the ratios and the first equality can be seen to lead also to eq.\,(\ref{a^2}).
Clearly the formulas in eqs.\,\,(\ref{|ASigma-|})-(\ref{ratios}) hold whether $CP$ symmetry is conserved or not.
It is worth remarking that implementing eq.\,(\ref{a^2}) in experimental fits involving $\alpha$ and $\beta$ (or $\phi$) of the $\mathit\Sigma^-$ and $\,\overline{\!\mathit\Sigma}{}^+$ modes could help achieve improved precision.

From the foregoing, we further learn that \,$\mathit\Sigma^-\to n\pi^-$\, and $~\overline{\!\mathit\Sigma}{}^+\to\overline n\pi^+$\, together are physically characterized by merely four real constants.
They correspond to four observables which may be chosen to be the rate
\,$\Gamma_{\mathit\Sigma^-\to n\pi^-}=\Gamma\raisebox{1pt}{$_{\,\overline{\!\mathit\Sigma}{}^+\to\overline n\pi^+}$}$\, and three of the parameters $\alpha_{\mathit\Sigma^-\to n\pi^-}$, $\alpha_{\,\overline{\!\mathit\Sigma}{}^+\to\overline n\pi^+}$, $\beta_{\mathit\Sigma^-\to n\pi^-}$ $\big({\rm or}~\phi_{\mathit\Sigma^-\to n\pi^-}\big)$, and $\beta_{\,\overline{\!\mathit\Sigma}{}^+\to\overline n\pi^+}$ $\big({\rm or}~\phi_{\,\overline{\!\mathit\Sigma}{}^+\to\overline n\pi^+}\big)$.

\section{Enhanced \textit{CP}-violation in \boldmath$\mathit\Lambda\to N\pi$ and $\mathit\Xi\to\mathit\Lambda\pi$ due to CMOs\label{LXcpv-cmo}}

The lowest-order effects of $\mathfrak L_\chi^{\tt CMO}$ in eq.\,(\ref{Lgnew}) on these channels are represented by diagrams analogous to the ones in figure~\ref{npdiagrams}.
Accordingly, we have~\cite{Tandean:2003fr}
\begin{align} \label{L->Npi}
\mathbb A_{\mathit\Lambda\to p\pi^-}^{\tt CMO}\, & =\, \frac{\beta_D^-+3\beta_F^-}{2\sqrt3\,f_\pi^{}} + \frac{\sqrt3\, \beta_\varphi^-}{2 f_\pi^{}} \bigg( \frac{m_N^{}-m_\mathit\Lambda^{}}{\hat m-m_s^{}} \bigg) \,=\, -\sqrt2\, \mathbb A_{\mathit\Lambda\to n\pi^0}^{\tt CMO} \,,
\nonumber \\
\mathbb A_{\mathit\Xi^-\to\mathit\Lambda\pi^-}^{\tt CMO}\, & =\, \frac{\beta_D^--3\beta_F^-}{2\sqrt3\,f_\pi^{}} + \frac{\sqrt3\, \beta_\varphi^-}{2f_\pi^{}} \bigg( \frac{m_\mathit\Xi^{}-m_\mathit\Lambda^{}}{\hat m-m_s^{}} \bigg) \,=\, -\sqrt2\, \mathbb A_{\mathit\Xi^0\to\mathit\Lambda\pi^0}^{\tt CMO} \,, &
\end{align}
\begin{align}
\mathbb B_{\mathit\Lambda\to p\pi^-}^{\tt CMO}\, & =\, \frac{m_\mathit\Lambda^{}+m_N^{}}{2\sqrt3\,f_\pi^{}} \Bigg[ ({\cal D+F}) \frac{\beta_D^++3\beta_F^+}{m_\mathit\Lambda^{}-m_N^{}} + 2{\cal D}\, \frac{\beta_D^+-\beta_F^+}{m_\mathit\Sigma^{}-m_N^{}} + \frac{{\cal D}+3\cal F}{m_s^{}-\hat m}\, \beta_\varphi^+ \Bigg]
\nonumber \\ & =\, -\sqrt2~ \mathbb B_{\mathit\Lambda\to n\pi^0}^{\tt CMO} \,,
\nonumber \\
\mathbb B_{\mathit\Xi^-\to\mathit\Lambda\pi^-}^{\tt CMO}\, & =\, \frac{m_\mathit\Lambda^{}+m_\mathit\Xi^{}}{2\sqrt3\,f_\pi^{}} \Bigg[ ({\cal D-F}) \frac{\beta_D^+-3\beta_F^+}{m_\mathit\Lambda^{}-m_\mathit\Xi^{}} + 2 {\cal D}\, \frac{\beta_D^++\beta_F^+}{m_\mathit\Sigma^{}-m_\mathit\Xi^{}} + \frac{{\cal D}-3\cal F}{m_s^{}-\hat m}\, \beta_\varphi^+ \Bigg] ~~~ ~~~~
\nonumber \\ & =\, -\sqrt2~ \mathbb B_{\mathit\Xi^0\to\mathit\Lambda\pi^0}^{\tt CMO} \,.
\end{align}
The CMO contributions to the weak phases in the \,$\Delta I=1/2$\, amplitudes for these $\mathit\Lambda$ and $\mathit\Xi$ modes can then be evaluated.
The outcomes are
\begin{align} \label{xiL-cmo}
\xi_{1\mathit\Lambda}^{\rm S,\tt CMO} & \,=\, -2.3\times10^5 {\rm~GeV~Im}\,\textsl{\texttt C}_g^- \,, & \xi_{1\mathit\Lambda}^{\rm P,\tt CMO} & \,=\, -2.4\times10^5 {\rm~GeV~Im}\,\textsl{\texttt C}_g^+ \,, ~~~ ~~~~ \nonumber \\
\xi_{1\mathit\Xi}^{\rm S,\tt CMO} & \,=\, -1.9\times10^5 {\rm~GeV~Im}\,\textsl{\texttt C}_g^- \,, & \xi_{1\mathit\Xi}^{\rm P,\tt CMO} & \,=\, 1.2\times10^5 {\rm~GeV~Im}\,\textsl{\texttt C}_g^+ \,,
\end{align}
which are expected to be only accurate up to relative uncertainties of order unity, as in the \,$\mathit\Sigma\to N\pi$\, case.

In estimating the $CP$ asymmetries, we again take into account the bounds in eqs.\,\,(\ref{ImCg+bound}) and~(\ref{ImCg-bound}).
The pertinent strong phases are \,$\delta_{1\mathit\Lambda}^{\rm S} = 6.52^\circ \pm 0.09^\circ$,\, $\delta_{3\mathit\Lambda}^{\rm S} = -4.60^\circ \pm 0.07^\circ$,\, $
\delta_{1\mathit\Lambda}^{\rm P} = -0.79^\circ \pm 0.08^\circ$,\, and \,$\delta_{3\mathit\Lambda}^{\rm P} = -0.75^\circ \pm 0.04^\circ$\, for \,$\mathit\Lambda\to N\pi$\,  \cite{Salone:2022lpt,Hoferichter:2015hva} and \,$\delta_\mathit\Xi^{\rm P}-\delta_\mathit\Xi^{\rm S} = 1.7^\circ \pm 1.1^\circ$\, for \,$\mathit\Xi\to\mathit\Lambda\pi$\, \cite{ParticleDataGroup:2024cfk,He:2025jfc}.
We also employ the empirical isospin amplitudes extracted in ref.\,\cite{He:2025jfc}.

Implementing steps in like manner to those applied for reaching eq.\,(\ref{cpa-cmo}) then leads to the averaged asymmetries and their 1$\sigma$ errors
\begin{align} \label{cpa-L-cmo}
\hat A_{\mathit\Lambda\to p\pi^-}^{\tt CMO} & \,=\, (-1.2\pm4.3) \times10^{-4} \,, & \hat A_{\mathit\Lambda\to n\pi^0}^{\tt CMO} & \,=\, (-1.0\pm4.0) \times10^{-4} \,,
\nonumber \\
\hat B_{\mathit\Lambda\to p\pi^-}^{\tt CMO} & \,=\, (-0.9\pm3.2) \times10^{-3} \,, & \hat B_{\mathit\Lambda\to n\pi^0}^{\tt CMO} & \,=\, (-0.8\pm3.4) \times10^{-3} \,,
\nonumber \\
\Delta\phi_{\mathit\Lambda\to p\pi^-}^{\tt CMO} & \,=\, (-1.0\pm3.6) \times10^{-3} \,, & \Delta\phi_{\mathit\Lambda\to n\pi^0}^{\tt CMO} & \,=\, (-0.7\pm3.0) \times10^{-3} \,,
\nonumber \\
\hat\Delta_{\mathit\Lambda\to p\pi^-}^{\tt CMO} & \,=\, (0.6\pm1.5) \times10^{-5} \,, & \hat\Delta_{\mathit\Lambda\to n\pi^0}^{\tt CMO} & \,=\, (-1.1\pm2.7) \times10^{-5} \,, ~~~ ~~
\end{align}
\begin{align} \label{cpa-X-cmo}
\hat A_{\mathit\Xi^-\to\mathit\Lambda\pi^-}^{\tt CMO} & \,=\, (2.2\pm6.5) \times10^{-5} \,, & \hat A_{\mathit\Xi^0\to\mathit\Lambda\pi^0}^{\tt CMO} & \,=\, (2.0\pm6.1) \times10^{-5} \,, ~~~ ~~
\nonumber \\
\hat B_{\mathit\Xi^-\to\mathit\Lambda\pi^-}^{\tt CMO} & \,=\, (-0.7\pm2.2) \times10^{-3} \,, & \hat B_{\mathit\Xi^0\to\mathit\Lambda\pi^0}^{\tt CMO} & \,=\, (-0.7\pm2.1) \times10^{-3} \,, ~~~ ~~
\nonumber \\
\Delta\phi_{\mathit\Xi^-\to\mathit\Lambda\pi^-}^{\tt CMO} & \,=\, (3.0\pm8.8) \times10^{-4} \,, & \Delta\phi_{\mathit\Xi^0\to\mathit\Lambda\pi^0}^{\tt CMO} & \,=\, (2.5\pm7.7) \times10^{-4} \,,
\end{align}
\begin{align} \label{cpa-LX-cmo}
\hat A_{\mathit\Lambda\to p\pi^-}^{\tt CMO} {+} \hat A_{\mathit\Xi^-\to\mathit\Lambda\pi^-}^{\tt CMO} & \,=\, (-1.0{\pm}4.4) \!\times\!10^{-4} , & A_{\mathit\Lambda\to n\pi^0}^{\tt CMO} {+} \hat A & _{\mathit\Xi^0\to\mathit\Lambda\pi^0}^{\tt CMO} \,=\, (-0.7{\pm}4.1) \!\times\!10^{-4} ,
\end{align}
as well as \,$\xi_{1\mathit\Lambda}^{\rm P,\tt CMO}-\xi_{1\mathit\Lambda}^{\rm S,\tt CMO} = (-0.8\pm3.3) \times10^{-3}$\, and \,$\xi_{1\mathit\Xi}^{\rm P,\tt CMO}-\xi_{1\mathit\Xi}^{\rm S,\tt CMO} = (-0.7\pm2.1) \times10^{-3}$.\,
Since the total isospin of the final states of \,$\mathit\Xi\to\mathit\Lambda\pi$\, has just one value, \,$I_{\rm f}=1$,\, their rate asymmetries vanish, \,$\hat\Delta_{\mathit\Xi^-\to\mathit\Lambda\pi^-} = \hat\Delta_{\mathit\Xi^0\to\mathit\Lambda\pi^0} = 0$,\, like \,$\mathit\Sigma^-\to n\pi^-$.\,
The $\hat A$ and $\hat B$ results in eq.\,(\ref{cpa-L-cmo})~[(\ref{cpa-X-cmo})], plus the weak phases, satisfy the approximations \,$\hat A_{\mathit\Lambda\to N\pi} = -{\rm tan}\big(\delta_{1\mathit\Lambda}^{\rm P}-\delta_{1\mathit\Lambda}^{\rm S}\big)\, {\rm tan}\big(\xi_{1\mathit\Lambda}^{\rm P}-\xi_{1\mathit\Lambda}^{\rm S}\big)$\,  and \,$\hat B_{\mathit\Lambda\to N\pi} = {\rm tan}\big(\xi_{1\mathit\Lambda}^{\rm P}-\xi_{1\mathit\Lambda}^{\rm S}\big)$\, [analogous approximations for \,$\mathit\Xi\to\mathit\Lambda\pi$]
valid to lowest order in the \,$\Delta I=3/2$\, amplitudes~\cite{Donoghue:1985ww}, and $\hat\Delta_{\mathit\Lambda\to p\pi^-,\mathit\Lambda\to n\pi^0}^{\tt CMO}$ are consistent with \,$2\hat\Delta_{\mathit\Lambda\to p\pi^-}\simeq-\hat\Delta_{\mathit\Lambda\to n\pi^0}$\, based on the rate data and $CPT$ theorem~\cite{Brown:1983wd}.

\begin{figure}[h] \bigskip \centering
\includegraphics[height=2.5in]{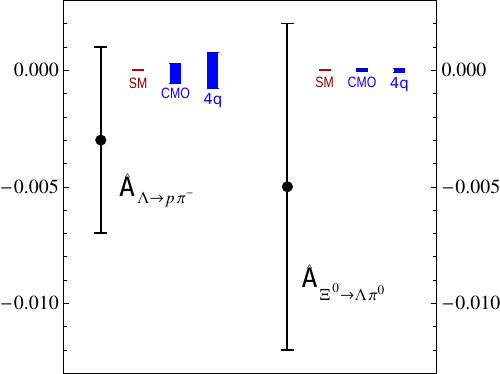} ~ ~ ~ \includegraphics[height=2.5in]{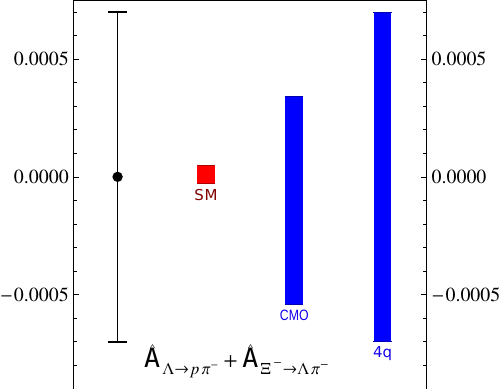}\vspace{-1pt}
\caption{The predicted $\hat A_{\mathit\Lambda\to p\pi^-}$ and $\hat A_{\mathit\Xi^0\to\mathit\Lambda\pi^0}$ (left panel) and \,$\hat A_{\mathit\Lambda\to p\pi^-}\!+\!A_{\mathit\Xi^-\to\mathit\Lambda\pi^-}$ (right panel) in the SM (red) and in the new-physics scenarios (blue) dealt with in sections~\ref{cmo} (\textsf{\small CMO}) and~\ref{dmmodel} (\textsf{\small 4q}), compared to the 1$\sigma$ intervals of the corresponding data cited in eq.\,(\ref{AL-AX-data}).} \label{AL-AX-exp-th}
\end{figure}

The $\hat A_{\mathit\Lambda\to N\pi}^{\tt CMO}$ and $\hat A_{\mathit\Xi\to\mathit\Lambda\pi}^{\tt CMO}$ ranges above exceed their SM expectations~\cite{He:2025jfc} by roughly an order of magnitude, but are less than the corresponding CMO results of ref.\,\cite{Tandean:2003fr} because of a stricter $\varepsilon$ constraint and smaller \,$\delta_\mathit\Xi^{\rm P}-\delta_\mathit\Xi^{\rm S}$.\,
Several of the predictions in eqs.\,\,(\ref{cpa-L-cmo})-(\ref{cpa-LX-cmo}) can also be compared to the most recent  data~\cite{ParticleDataGroup:2024cfk}:
\begin{align} \label{AL-AX-data}
\hat A_{\mathit\Lambda\to p\pi^-}^{\rm exp} & \,=\, (-3\pm4)\times10^{-3} \,, &
A_{\mathit\Lambda\to n\pi^0}^{\rm exp} & \,=\, \big(1_{-11}^{+10}\big)\times10^{-3} \mbox{\, \cite{BESIII:2023jhj}} \,,
\nonumber \\
\hat A_{\mathit\Xi^-\to\mathit\Lambda\pi^-}^{\rm exp} & \,=\, \big({-}9_{-\,\,8}^{+11}\big)\times10^{-3} \mbox{\, \cite{BESIII:2023jhj}} \,, & \hat A_{\mathit\Xi^0\to\mathit\Lambda\pi^0}^{\rm exp} & \,=\, (-5\pm7)\times10^{-3} \mbox{\, \cite{BESIII:2023drj}} \,, ~~~~
\nonumber \\
\Delta\phi_{\mathit\Xi^-\to\mathit\Lambda\pi^-}^{\rm exp} & \,=\, \big({-}3_{-11}^{+\,\,9}\big)\times10^{-3} \mbox{\, \cite{BESIII:2023jhj}} \,, & \Delta\phi_{\mathit\Xi^0\to\mathit\Lambda\pi^0}^{\rm exp} & \,=\, (0\pm7)\times10^{-3} \mbox{\, \cite{BESIII:2023drj}} \,,
\nonumber \\
\big(\hat A_{\mathit\Lambda\to p\pi^-} + \hat A & _{\mathit\Xi^-\to\mathit\Lambda\pi^-}\big){}_{\rm exp}^{} \,=\, (0\pm7)\times10^{-4} \mbox{\, \cite{HyperCP:2004zvh}} \,.
\end{align}
In figure \ref{AL-AX-exp-th}, we display the 1$\sigma$ ranges of $\hat A_{\mathit\Lambda\to N\pi,\mathit\Xi\to\mathit\Lambda\pi}$ that have the strictest experimental limits. Evidently, at the moment $\big(\hat A_{\mathit\Lambda\to p\pi^-} \!+\! \hat A_{\mathit\Xi^-\to\mathit\Lambda\pi^-}\big){}_{\rm exp}^{}$ alone is close to probing its CMO counterpart.
In the THDM+D discussed in section~\ref{dmmodel}, the situation is different in that the empirical bound on this asymmetry sum can be saturated, as indicated on the second panel (blue-colored band labeled \textsf{4q}) of this figure and elaborated in ref.\,\cite{He:2025jfc}.

\bibliographystyle{utphys.bst}
\bibliography{refs.bib}

\end{document}